# Comparison analysis between standard polysomnographic data and in-ear-EEG signals: A preliminary study

Gianpaolo Palo[1,2,†], Luigi Fiorillo[1,*, †], Giuliana Monachino[1,3], Michal Bechny[1,3], Michel Wälti[4], Elias Meier[4], Francesca Pentimalli Biscaretti di Ruffia[4], Mark Melnykowycz[4], Athina Tzovara[3], Valentina Agostini[2], and Francesca Dalia Faraci[1]

[1]Institute of Digital Technologies for Personalized Healthcare (MeDiTech), Department of Innovative Technologies, University of Applied Sciences and Arts of Southern Switzerland, Lugano, Switzerland; [2]Department of Electronics and Telecommunications, Politecnico di Torino, Torino, Italy; [3]Institute of Computer Science, University of Bern, Bern, Switzerland; [4]IDUN Technologies AG, Glattpark, Switzerland.

Institution where work was performed: Institute of Digital Technologies for Personalized Healthcare (MeDiTech), Department of Innovative Technologies, University of Applied Sciences and Arts of Southern Switzerland, Lugano, Switzerland.

†These authors contributed equally to this work.

*Corresponding author. Luigi Fiorillo, Institute of Digital Technologies for Personalized Healthcare (MeDiTech), Department of Innovative Technologies, University of Applied Sciences and Arts of Southern Switzerland, Lugano, Switzerland. Email: luigi.fiorillo@supsi.ch.

# Abstract

**Study Objectives:** Polysomnography (PSG) currently serves as the benchmark for evaluating sleep disorders. Its discomfort makes long-term monitoring unfeasible, leading to bias in sleep quality assessment. Hence, less invasive, cost-effective, and portable alternatives need to be explored. One promising contender is the in-ear-EEG sensor. This study aims to establish a methodology to assess the similarity between the single-channel in-ear-EEG and standard PSG derivations.

**Methods:** The study involves four-hour signals recorded from ten healthy subjects aged 18 to 60 years. Recordings are analyzed following two complementary approaches: (i) a hypnogram-based analysis aimed at assessing the agreement between PSG and in-ear-EEG-derived hypnograms; and (ii) a feature-based analysis based on time- and frequency- domain feature extraction, unsupervised feature selection, and definition of Feature-based Similarity Index via Jensen-Shannon Divergence (JSD-FSI).

**Results:** We find large variability between PSG and in-ear-EEG hypnograms scored by the same sleep expert according to Cohen's kappa metric, with significantly greater agreements for PSG scorers than for in-ear-EEG scorers ($p < 0.001$) based on Fleiss' kappa metric. On average, we demonstrate a high similarity between PSG and in-ear-EEG signals in terms of JSD-FSI ($0.79 \pm 0.06$ -awake, $0.77 \pm 0.07$ -NREM, and $0.67 \pm 0.10$ -REM) and in line with the similarity values computed independently on standard PSG-channel-combinations.

**Conclusions:** In-ear-EEG is a valuable solution for home-based sleep monitoring, however further studies with a larger and more heterogeneous dataset are needed.

**Keywords:** sleep wearables, in-ear-EEG, machine learning, sleep staging, multi-source-scored sleep databases.

## Statement of Significance

Traditional polysomnography may prevent from depicting real sleep patterns due to the extensive setting employed. An alternative to overcome this limitation is to use wearable solutions like the in-ear-EEG. To date, the in-ear-EEG and the standard PSG derivations have only been compared following basic correlation analysis. We propose a more exhaustive methodology - hypnogram-based and feature-based - to evaluate the similarity between the in-ear-EEG and PSG signals. The ultimate goal is to investigate whether in-ear-EEG sensors inherit information close to the ones we extract through standard PSG.

# Introduction

Sleep is essential to good health. Poor or inadequate sleep is associated with several dysfunctions in most physiological systems. Sleep analysis is of crucial importance in the diagnosis and treatment of sleep disorders [1, 2].

Polysomnography (PSG) is the gold standard to perform sleep studies [1-3]. PSG is performed in appropriate clinical facilities, and involves recording multiple bio-signals during a full night's sleep, including brain activity (EEG), eye movements (EOG), muscle activity (EMG), cardiac activity (ECG), body position, breathing effort, blood saturation, etc [1, 2]. The PSG recordings are nowadays manually evaluated by trained personnel according to the American Academy of Sleep Medicine (AASM) manual [4]. Despite being highly standardized by AASM guidelines, this manual procedure is time- and effort- consuming, and it is not error-free [5]. These limitations, along with the saturation of the sleep units, lead to high costs related to patient management and care. Besides, due to the invasive equipment, and since the patients are typically sleeping in an atypical and unfamiliar environment, standard PSG-based analyses introduce biases to the sleep quality assessment [1, 2].

Wearable and portable devices may be valid solutions, as they allow for home-based sleep monitoring. The use of unconventional channels has been widely explored in the field of mobile sleep monitoring with wearable devices. A comprehensive overview about sensing technologies (different signals and their combinations) for sleep staging via wearable devices is provided in [6]. The signals conveying a substantial amount of information for this task are EEG, EOG, and EMG - with the EEG signal being the most used sensing modality as a single data source. We might therefore speculate that ear-EEG may be the right choice. The brain activity is recorded from electrodes placed in or around the ear, while also leading to several advantages in comfort, fixed electrode positions, robustness to electromagnetic interference, and ease of use [1, 7].

To date, only two research groups [2, 8-13] have exploited ear-EEG signals for sleep analysis. The majority of their studies mainly relied on feature-based methodologies to evaluate the feasibility of ear-EEG technology for automated sleep monitoring. They showed that automatic sleep scoring based on ear-EEG signals was performing at levels comparable to expert scoring of PSG, in young healthy subjects [2, 8-13].

Jørgensen et al. [14] study can be seen as a proof of concept for the suitability of ear-EEG on epileptic subjects and in [15], Kjaer et al. showed that sleep metrics computed from multiple nights automatically scored on ear-EEG are more reliable than the ones computed from a single night manually scored via standard PSG. Thus, highlighting how ear-EEG seems to be a useful alternative for sleep staging for the single night recording, and an advantageous choice for several nights of sleep monitoring.

However, in none of the above mentioned studies - even before inferring and/or validating sleep metrics and/or algorithms on these promising signals - the similarity between each standard PSG and the ear-EEG derivations has been thoroughly investigated or quantified.

In this work, we carried out the above-mentioned comparison analysis, first focusing on the sleep scoring procedure (*hypnogram-based*), and then directly evaluating the signals (*feature-based*). In the *Methods Section*, we briefly describe the dataset along with the instrumentation, data collection procedure, and pre-processing of the signals. In the *Comparison analysis: hypnogram-based approach sub-section,* we first describe how to define the consensus in a multi-source-scored dataset. The hypnogram-based comparison analysis is performed by evaluating the intra- and inter- scorer variability, thus assessing the agreement (i.e., Cohen's kappa and Fleiss' kappa) between the PSG and in-ear-EEG derived hypnograms. In the *Comparison analysis: feature-based approach sub-section* we present all the steps of our feature-based comparison analysis, i.e., time- and frequency- domain feature extraction, feature selection, and the final evaluation of the similarity between the two different sources. The proposed approach relies on a comparison of the distributions of the selected features - extracted from the two different sources, PSG and in-ear-EEG respectively - via the newly introduced Jensen–Shannon Divergence Feature-based Similarity Index (JSD-FSI). In-ear-EEG earbuds sensors are thought to perform better - for sleep scoring tasks - when combined with additional EOG signals [3, 10, 16] - especially in distinguishing the REM sleep stage. Therefore, we extract features from both EOG and scalp-EEG derivations (i.e., frontal, central, and occipital brain regions), and we compare them to the features extracted from the in-ear-EEG recordings. Among the scalp-EEG derivations, we also included the mastoid-to-mastoid one (M1-M2), as its information has been proven to be similar to the in-ear-EEG [7, 11].

In the *Results Section*, we present the most significant outcomes of our hypnogram-based and feature-based approaches, validating any related observations through appropriate statistical analyses. Finally, in the *Conclusions Section,* we discuss the main key points and implications of our findings, highlighting the contributions of the current study as well as the limitations encountered.

To summarize, in this work we investigate whether or not in-ear-EEG sensors inherit information - or set of features - close to what we usually extrapolate through standard PSG derivations.

The main contributions of our study are the following:

1. we found large intra-scorer variability related to the in-ear-EEG scoring compared to the PSG scoring, with agreements among PSG scorers significantly ($p < 0.001$) greater than the ones among in-ear-EEG scorers. This difference is probably due to the uncertainty the scorers have when evaluating in-ear-EEG signals;

2. we show that the similarity between the PSG and the in-ear-EEG signals - in terms of JSD-FSI score - is high, on average $0.79 \pm 0.06$ in awake, $0.77 \pm 0.07$ in NREM and $0.67 \pm 0.10$ in REM;

3. we found significant changes in JSD-FSI scores between sleep stages, with significantly greater values for the awake stage with respect to NREM ($p < 0.001$), and REM (p < 0.001) stages, and significantly greater values for the NREM stage compared to REM (p < 0.001) stage;
4. we prove that the JSD-FSI similarity values reported in (2) are revealed to be in line/overlap with the similarity values computed independently on the different combinations of PSG channels.

## Methods

We exploit an already existing dataset collected during an observational study carried out at IDUN Technologies. In the hypnogram-based comparison analysis, we first describe how to compute the consensus in the multi-source-scored dataset (i.e., a dataset where each recording is scored by multiple experts and looking at different sources of signals), and then we assess the intra- and inter- scorer variability. Establishing a consensus is crucial to better conduct the feature-based comparison analysis - i.e., analyzing only the sleep epochs where the PSG-scorers and the in-ear-EEG-scorers were in agreement on the associated sleep period.

Indeed, in our feature-based comparison procedure we evaluate the similarity between the signals coming from two different sources for each sleep stage independently. The feature-based analysis is divided in three steps: feature extraction (time- and frequency-domain features), feature selection, and calculation of the newly defined Jensen-Shannon Divergence (JSD) - Feature-based Similarity Index (FSI).

### Dataset

The quality assurance study (BASEC Nr. Req-2022-00105) involves 10 healthy subjects, including both females and males (18-60 years) selected according to the Pittsburgh sleep quality index (PSQI) [17]. Following a screening period of 28 days, the subjects experience one overnight stay at the investigational site. Participants are monitored using multiple standard surface electrodes on their scalp (EEG), outer canthus of each eye (EOG), mentalis (chin), torso (EKG) in the conventional PSG monitoring, and an additional in-ear-EEG sensing technology monitoring. Participants arrive approximately three hours prior to their normal bedtime at the sleep laboratory, and they are instructed about the overall study, including PSG preparation /setting phase, a sleep restriction phase ≈ 4 hours, and the sleep phase ≈ 4 hours ([Figure 1](#)).

The subjects are asked to avoid caffeine intake from noon before arriving at the study center. They perform specific head and eye movements for bio-calibrating the instrumentation used for data collection. Ear tips of three different sizes are given to participants to provide proper fit and electrical contact. A 10-minute stabilization period is awaited before starting data collection to ensure the in-ear electrodes reach thermal equilibrium with the participant's body temperature. For each participant, an impedance measurement of the in-ear-EEG signal is performed at 31.2 Hz to guarantee that the system is stable. A 300 kOhm threshold is set for good signal quality - greater values indicate either electrical malfunctions or incorrect placement of the ear tips in the ear.

The sleep restriction starts at the in-bed time of each subject and lasts for about four hours, after which participants are allowed to sleep for another four hours. During the last hour of the sleep restriction phase, the subjects abstain from using electronic devices. The sleep room is set up according to AASM guidelines [4]. The Karolinska Sleepiness Scale (KSS) [18] is administered at the beginning and at the end of the sleep restriction period, and prior to sleep to assess subjective drowsiness.

PSG and in-ear-EEG signals are recorded simultaneously. The data analyzed in this study refer only to the four hours recorded during sleep.

PSG signals are collected using a SOMNOmedics SOMNOscreen plus system with a sampling frequency of 256 Hz ([Figure 2a](#)). The signals are band-pass filtered between 0.2 - 35 Hz and ECG artifacts are removed automatically by the recording device. A total of 21 channels are investigated, considering both bipolar and unipolar derivations: two reference electrodes (M1, M2); six EEG bipolar derivations (C3-M2, F3-M2, O1-M2, C4-M1, F4-M1, O2-M1); six EEG unipolar derivations (C3, C4, F3, F4, O1, O2); four EOG bipolar derivations (E1-M1, E1-M2, E2-M1, E2-M2); two EOG unipolar derivations (E1, E2); the mastoid-to-mastoid derivation (M2-M1). From here on, we will refer to the set of PSG channels as the set $Q$.

In-ear-EEG signals are collected via the GDK (Guardian Development Kit) device designed by IDUN Technology with a sampling frequency of 250 Hz (Figure 2b). The GDK system includes hardware (Brain Computer Interface) and streaming software (Neuro-Intelligence Platform). The former involves two dry contact electrodes designed by IDUN Technology, i.e., Dryode Ink electrodes, which are made of an elastomer material functionalized by an electrically conductive coating. The recording and reference channels are placed in the right and left ears respectively. Biopotential differences between electrodes are measured using the ADS1299-x amplifier (Texas Instruments, LLC, Dallas, Texas, United States). The in-ear-EEG signals are band-pass filtered between 0.5 - 35 Hz, before being normalized as their amplitude range would match the one of the simultaneously recorded PSG signals. In-ear-EEG recordings are multiplied with the standard deviation ratio of PSG and in-ear-EEG data. From here on, we will refer to the single in-ear-EEG channel as CH1 channel. PSG and in-ear-EEG signals are manually synchronized based on easily distinguishable artifacts in both data streams. They are then trimmed such that the recordings referring to the same subject share the same length.

Three scoring experts have independently scored both signals - first evaluating PSG and then in-ear-EEG data, according to AASM guidelines [4]. This results in three PSG hypnograms and three in-ear-EEG hypnograms, for each subject. The dataset contains the following annotations W, N1, N2, N3, REM, MOVEMENT and UNKNOWN, where the last two refer respectively to movement artifacts and to no sleep stage assigned. In this study, the three non-REM sleep stages are combined together under the label NREM, and all the epochs scored as MOVEMENT or UNKNOWN are not considered.

## Comparison analysis: hypnogram-based approach

### Consensus in a multi-source-scored dataset

In the hypnogram-based approach, we first compute the consensus among the three scorers on each data source - inspired by previous studies [19, 20] analyzing multi-scored databases. The majority vote from the scorers has been computed - i.e., we assign to each 30-second epoch the most voted sleep stage among the scorers. In case of ties, we compute the *soft-agreement* metric [20] to then consider the label from the most reliable scorer. The most reliable scorer is the one that is frequently in agreement with all the others. We then rank the reliability of each scorer, to finally define the most reliable scorer, for each subject.

We denote with $J$ the total number of scorers, with $j$ the scorer for which the soft-agreement metric is evaluated, and with $i$ all the other scorers. The one-hot encoded sleep stages given by the scorer $j$ are: $\hat{y}_j \in [0, 1]^{K \times T}$, i.e., 1 assigned for the scored stage and 0 for the other stages, $K$ is the number of classes, i.e., $K = 3$ sleep stages, and $T$ is the total number of epochs. The probabilistic consensus $\hat{z}_j$ among the $J - 1$ scorers ($j$ excluded) is computed using the following:

$$\hat{z}_j = \frac{\sum_{i=1}^{J} \hat{y}_i[t]}{\max \sum_{i=1}^{J} \hat{y}_i[t]} \quad \forall t; \quad i \neq j \quad (1)$$

where $t$ is the $t\text{-}th$ epoch of $T$ epochs and $\hat{z}_j \in [0, 1]^{K \times T}$, i.e., 1 is assigned to a stage if it matches the majority or if it is involved in a tie. The maximum function in the denominator is used to combine the contributions from multiple scorers while ensuring that the scale of values remains within the range [0, 1]. The $Soft\text{-}Agreement$ is then computed separately for each scorer across all the $T$ epochs as:

$$Soft\text{-}Agreement_j = \frac{1}{T} \sum_{t=0}^{T} \hat{z}_j[y_j] \qquad (2)$$

where $\hat{z}_j[y_j]$ denotes the probabilistic consensus of the sleep stage chosen by the scorer $j$ for the $t\text{-}th$ epoch. $Soft\text{-}Agreement_j \in [0, 1]$, where the zero value is assigned if the scorer $j$ systematically scores all the annotations incorrectly compared to the others, whilst 1 is assigned if the scorer $j$ is always involved in tie cases or in the majority vote. The $Soft\text{-}Agreement$ is computed for all the scorers, and the values are sorted from the highest - high reliability - to the lowest - low reliability.

The $Soft\text{-}Agreement$ is computed for each subject, i.e., the scorers are ranked accordingly, and in case of a tie the top-1 scorer will be the one used for that subject. In Supplementary Table S1 and Table S2 we report the $Soft\text{-}Agreement$ values computed on each of the three scorers, and for each subject, on the PSG and in-ear-EEG data sources respectively.

Intra- and inter- scorer variability are then assessed using Cohen's kappa and Fleiss' kappa metrics [21], respectively. The intra-scorer variability refers to the comparison between PSG and in-ear-EEG hypnograms scored by the same clinician, while the inter-scorer variability characterizes the agreement among scorers related to the same source, i.e., either PSG or in-ear-EEG scorers. According to Landis and Koch [14], Cohen's kappa values exceeding 0.80 suggests an almost-perfect agreement between scorers; a range of 0.61-0.80 indicates substantial agreement, whereas 0.41–0.60 implies moderate agreement. Fair agreement falls in the range of 0.21–0.40, and slight agreement occurs between 0.00 and 0.20. Fleiss' kappa values are interpreted in the same way [21].

In the feature-based comparison analysis, we define similarity-scores between the two different data sources, PSG and in-ear-EEG, exploiting a per-sleep-stage-based approach. Hence, we first define a common label-ground-truth reference for both types of signal - to prevent additional bias in our analysis. Such a reference is defined by all the epochs scored in the same sleep stage by both the consensus, i.e., PSG and in-ear-EEG scoring procedures. Therefore, for each subject, starting from the PSG and in-ear-EEG hypnograms, we first evaluate the consensus among the three expert scorers for the PSG and the in-ear-EEG respectively, and then we consider only the epochs where these two consensus are in agreement for our sleep stage-wise feature analysis. In [Table 1](Table 1) we report a summary of the total number and percentage of the epochs per sleep stage for both PSG and in-ear-EEG based scoring procedure, and the intersection (∩) computed on the labels coming from the two different sources.

## Comparison analysis: feature-based approach

In order to assess the similarity between standard PSG and in-ear-EEG signals we follow specific steps (as summarized in [Figure 3a](Figure 3a)): (1) we extract time- and frequency- domain features from the above-mentioned PSG derivations and the in-ear-EEG channel on each 30-second sleep epoch; (2) we remove all redundant features through pairwise assessments (feature selection procedure) to identify those conveying the same information; (3) we then define the JSD - Feature based Similarity Index (FSI) exploited to compare the distributions of the selected features, for each sleep stage and for each subject on both PSG derivations and the in-ear-EEG.

To fairly validate the results of our comparison analysis, we decided to also quantify the similarity between all the possible combinations of PSG derived signals, including scalp-EEG and EOG channels ([Figure 3b](Figure 3b)). The idea is to assess that the PSG-to-In-ear-EEG JSD-FSI similarity-scores (histograms in blue [Figure 3a](Figure 3a)) are on average close to those we derive from the standard PSG-to-PSG comparisons (histograms in red [Figure 3b](#)). The results are evaluated separately for each sleep stage and for each subject.

## Feature extraction

We extract both time- and frequency- domain features from 30-second epochs of signals. All the features depending on the amplitude are computed on signals normalized by their maxima, compensating for differences in magnitudes.

### Time-domain features

First, we simply compute standard descriptive statistics (e.g., mean, standard deviation, interquartile range, skewness, kurtosis, etc..), the maximum first derivative, and the number of zero-crossings. We then include entropy-based measures, specifically, the approximate entropy [22-24], the sample entropy [22-26], the Singular Value Decomposition (SVD) entropy [25, 26], and the permutation entropy [25, 27]. To assess the complexity of the signals we consider the Lempel-Ziv complexity [28] and the Detrended Fluctuation Analysis (DFA) exponent [29, 30]. In addition, we also include Hjorth parameters of activity, mobility, and complexity [31], the Katz, Higuchi [32], and Petrosian [33] fractal dimensions [34, 35] to further quantify the complexity or irregularity of the analyzed time series.

### Frequency-domain features

We first compute the Power Spectral Density (PSD) of each 30-second signal using the Welch's average periodogram method [36]. We choose a Hamming window of 5-second length with a 50% overlap, resulting in a frequency resolution of 0.2 Hz [30, 37]. We exploited the Hamming window to reduce the estimation variance, the side-lobe effect, and the spectral leakage phenomena [38]. The 5-second window length is set to be at least twice the lowest frequency of interest 0.5 Hz (i.e., the lower end of the EEG delta power band) [37]. Then, the median-average partially mitigates the influence of any noise/artifacts we have on our signals [30, 37].

Once the PSD has been computed, we first extract standard frequency-domain features, such as the spectral energy of the whole 30-second signal, and the relative spectral power on all the EEG frequency bands, i.e., delta (δ, 0.5–4 Hz), theta (θ, 4–8 Hz), alpha (α, 8–12 Hz), sigma (σ, 12–16 Hz), beta (β, 16–30 Hz), and gamma (γ, 30–35 Hz). We then include several ratio measures between the different frequency bands, i.e., δ/θ, δ/σ, δ/β, θ/α, δ/α, α/β, δ/(α + β), θ/(α + β), and δ/(α + β + θ).

In our frequency domain analysis, we include additional features assessing the spread, symmetry, tail behavior, shape, and complexity of each 30-second signal's spectrum distribution. Specifically, we compute the four central moments in statistics (i.e., mean, variance, skewness, and kurtosis), the spectral entropy and the Renyi entropy [39], the spectral centroid [40, 41], the spectral crest factor [42], the spectral flatness [40, 41], the spectral roll-off [43], and the spectral spread [41].

In Supplementary Analyses, we report the complete list of all the time- and frequency-domain features extracted (Table S3 and Table S4) and additional mathematical details for each of the above features.

**Feature selection**

The feature selection procedure is essential to remove in our analysis possible redundancy within the feature subset. We exploit a feature selection algorithm based on pairwise feature correlation [44, 45], aiming to identify the most representative features among all the extracted ones. However, before proceeding with this procedure, we should first consider that the above derived features are meant to describe the morphology of our neurological signals. The features are all supposed to change based on the state brain subjects are in, i.e., waking state, NREM state, or REM state. Thus, we decide to first divide the data, i.e., the 30-second epochs, depending on the sleep stage they are assigned to.

Therefore, for each pair of channels (i.e., a pair is defined as $\{q, CH1\}$, where $q \in Q$, and $Q$ is the above defined set of PSG channels), we build pairs of datasets $\{D^q, D^{CH1}\}$, one pair for each of the $k \in K$ sleep stages. Each dataset $D \in \Re^{MxNxK}$ is the result of concatenated feature vectors $\overline{f}_{n,k}$, where $K = 3$ is the number of sleep stages, $N$ is the total number of features, with $n \in N$, and $M$ is the total number of 30-second epochs in each stage.

A z-score normalization is performed separately on each dataset-pair, $D^q$ and $D^{CH1}$, and for each sleep stage, to reduce dissimilarities among the different subjects. On each dataset, we perform the feature selection based on the computation of a modified version of the maximal information compression index (MICI) between each pair of features [44-46]. As the algorithm adopts a k-nearest neighbors approach, the determination of the initial k value is crucial and is guided by metrics such as the representation entropy [44, 45] and the redundancy rate [45, 47].

In Supplementary Analyses, we report additional mathematical details regarding the modified version of the feature selection algorithm, along with further details on the k-nearest neighbors approach.

**JSD - Feature based Similarity Index (JSD-FSI)**

We quantify the similarity between pairs of feature distributions, coming from two different sources (e.g., PSG derived and in-ear-EEG derived), exploiting the Jensen-Shannon Divergence (JSD) [48]. The JSD divergence is a symmetric and smoothed version of the Kullback-Leibler (KL) divergence [49], quantifying the similarity between two probability distributions. Practically, we first compute the probability density function (PDF) $\Phi$ for each pair of feature distributions $\{\Phi(\overline{f}_{n,k}^{q}), \Phi(\overline{f}_{n,k}^{CH1})\}$ extracted from the paired datasets $\{D^q, D^{CH1}\}$ [50]. We then measure the dissimilarities between each pair via the JSD divergence. JSD ranges from 0 (identical distributions) to 1 (completely dissimilar distributions). The higher the number, the more dissimilar the probability distributions. Hence, once we compute the JSD metric on each PDF feature pair $\{\Phi(\bar{f}_{n,k}^q), \Phi(\bar{f}_{n,k}^{CH1})\}$, for each sleep stage $k$ and of each subject, we can finally compute the JSD Feature based Similarity Index ($JSD\text{-}FSI$) between each PSG $q$ channel and the in-ear-EEG $CH1$ channel as follows:

$$JSD\text{-}FSI_{k \in K} = \sum_{n=1}^{N} (1 - JSD_n) / N \qquad (3)$$

$$JSD_n(\Phi(\bar{f}_{n,k}^q) \| \Phi(\bar{f}_{n,k}^{CH1})) = \left(KL\left(\Phi(\bar{f}_{n,k}^q) \| M\right) + KL\left(\Phi(\bar{f}_{n,k}^{CH1}) \| M\right)\right) / 2 \qquad (4)$$

where $N$ is the total number of features, $M$ is the *average* distribution defined as $M = \left(\Phi(\bar{f}_{n,k}^q) + \Phi(\bar{f}_{n,k}^{CH1})\right)/2$, whilst $KL\left(\Phi(\bar{f}_{n,k}^q) \| M\right)$ is the Kullback-Leibler divergence between the two distributions $\Phi(\bar{f}_{n,k}^q)$ and $M$, defined as:

$$KL(\Phi(\bar{f}_{n,k}^q) \| M) = \sum f_k^q(t) * log\left(f_k^q(t) / M(t)\right) \quad \forall t; \qquad (5)$$

Hence, for each sleep stage and for each subject, we derive 21 (i.e., total number of pairs comparisons $\{q, CH1\}$) $JSD\text{-}FSI$ similarity-scores. Each of these scores is defined as the sum of the individual $JSD_n$ similarity-scores - resulting from all the PDF feature distributions comparisons - divided by the total number of features analyzed *(2)*.

The same comparison analysis has been done between all the possible combinations of PSG derived signals, i.e., scalp-EEG-to-scalp-EEG, scalp-EEG-to-EOG, and EOG-to-EOG comparisons. We compare all the PDF feature distributions extracted from all the channels in the PSG set $Q$ (unique comparisons, i.e., upper triangle of the symmetric matrix with dimension $(|Q| \times |Q|)$, where $|Q| = 21$ is the cardinality of the set, or the total number of PSG channels). Hence, we first compute the probability density function (PDF) $\Phi$ for each pair of feature distributions $\{\Phi(\overline{f}_{n,k}^{q_i}), \Phi(\overline{f}_{n,k}^{q_j})\}$ extracted from the dataset pairs $\{D^{q_i}, D^{q_j}\}$, with $i \neq j$. We then compute the $JSD\text{-}FSI$ similarity-scores on all the possible combinations as described above. In that case, for each sleep stage and for each subject, we derive 210 (i.e., unique comparisons between all the PSG channels via the binomial coefficient $\frac{|Q|!}{2!(|Q|-2)!}$ ) $JSD\text{-}FSI$ similarity-scores.

We will fairly assess, for each sleep stage and for each subject, that the PSG-to-In-ear-EEG JSD-FSI similarity-scores are on average close to those derived from the standard PSG-to-PSG comparisons.

## Results

### Comparison analysis: hypnogram-based approach

**Intra- and inter- scorer variability in the multi-source-scored dataset**

We measure the agreement between each pair of PSG and in-ear-EEG hypnograms referring to the same recording/subject scored by the same scorer expert, according to Cohen's kappa and Fleiss' kappa metrics.

In [Figure 4](#) we report, for each scorer, the distribution of the Cohen's kappa values computed for each recording/subject between the PSG and in-ear-EEG hypnograms - so quantifying the intra-scorer variability in the multi-source-scored dataset. In this context, the values exhibit considerable dispersion across all distributions, with limited agreement levels, particularly in the comparison performed on the scorer 2. No significant changes are found among the three Cohen's kappa distributions according to the ANOVA test ($\alpha = 0.05$). The normality assumption is verified based on the Shapiro-Wilk test ($\alpha = 0.05$).

In [Figure 5](#) we report, for each data source, the distribution of the Fleiss' kappa values, comparing hypnograms from the same recording/subject, scored by the three expert scorers, first on PSG and then on in-ear-EEG signals - so quantifying the inter-scorer variability in the multi-source-scored dataset.

Notably, the three PSG scorers exhibit greater coherence scoring the PSG recordings, compared to when scoring the in-ear-EEG signals. Indeed, Fleiss' kappa values for PSG hypnograms are found to be significantly greater (p < 0.001) than those between in-ear-EEG hypnograms, based on the Student's t-test ($\alpha = 0.05$). We employ a parametric statistical test given normal distributions, as stated by the Shapiro-Wilk test ($\alpha = 0.05$).

For each sleep stage, we also assess the average agreement across all subjects between the PSG and in-ear-EEG consensus ([Table 2](#)), finding large discrepancies for the REM stage. The agreement is evaluated according to accuracy, precision, recall, and F1-score metrics, using the PSG consensus as the gold standard.

## Comparison analysis: feature-based

### JSD-FSI similarity-scores

The most frequently selected features across the different datasets $\{D^q, D^{CH1}\}$ for all the three sleep stages ([Figure 6](#)) are the following: the relative δ, θ, α, and σ power bands; the δ/θ power ratio; the spectral flatness; the spectral variance; the skewness; the kurtosis; the maximum first derivative; and the Hjorth activity and complexity. In addition, there are extra selected features specifically for each sleep stage: the spectral skewness, the inter-quartile range, the Hjorth mobility, the Renyi entropy, the permutation entropy, and the Higuchi fractal dimension for the awake stage; the relative β and γ power bands, the spectral skewness, the standard deviation, the number of zero-crossings, the spectral entropy and the permutation entropy for the NREM sleep stage; and the spectral energy, the relative γ power band, the spectral kurtosis, the standard deviation, and the spectral entropy for the REM sleep stage. The above selected features are not to be understood as more or less relevant to the purpose of our comparison analysis. The not-selected features were ignored because of the redundant information they were bringing.

The common label-ground-truth reference (i.e., intersection computed on the labels coming from the two different sources) for subjects 3 and 8 do not show any REM epochs - lack of agreement between corresponding PSG and in-ear-EEG consensus. Therefore, the results will not include JSD-FSI similarity scores for these two subjects in the REM class. Furthermore, whenever the in-ear-EEG channel or any of the PSG channels show some noisy epochs (e.g., no signal - constant amplitude - meaning no reliable indicators of brain activity), we exclude that specific channel. This situation occurs mainly for the channel M2 for subjects 3 and 6 - with 79% and 82% of noisy epochs respectively. Hence, we exclude the channel M2 from the analysis for subjects 3 and 6.

In [Figure 7](#), [Figure 8](#), and [Figure 9](#) we report the JSD-FSI similarity-scores computed between the in-ear-EEG and PSG channels - for each sleep stage respectively - in standard topographic images. Overall, the similarity between the PSG and the in-ear-EEG signals - in terms of JSD-FSI score - is high, on average 0.79 ± 0.06 in awake, 0.77 ± 0.07 in NREM, and 0.67 ± 0.10 in REM. On average, there are no substantial differences with the in-ear-EEG compared to the corresponding PSG channel across all subjects. The spatial distributions in terms of JSD-FSI scores (any pair in-ear-EEG and PSG derivations) are on average consistent within the different subjects and channels. According to the Kruskal-Wallis test ($\alpha = 0.05$), significant changes in JSD-FSI scores are found among sleep stages (p < 0.001). In detail, based on the Mann-Whitney U-test ($\alpha = 0.05$) there is statistical evidence that JSD-FSI similarity scores for the awake stage are greater than NREM (p < 0.001) and REM (p < 0.001) ones; and that JSD-FSI similarity scores for the NREM stage are greater than REM ones (p < 0.001). Non-parametric statistical tests are used as the normality assumption is not met based on the Shapiro-Wilk test ($\alpha = 0.05$).

When assessing the similarity between in-ear-EEG and PSG derivations, and comparing it to the similarity computed among all the possible 210 PSG-to-PSG comparisons (JSD-FSI similarity-scores computed from the PSG channels), similar values are observed.
In Figure 10, Figure 11, and Figure 12 we show that, for every sleep stage, the blue distributions (i.e., JSD-FSI similarity-scores from the PSG-to-In-ear-EEG comparisons) consistently align with the red ones (i.e., JSD-FSI similarity-scores from the PSG-to-PSG comparisons). The only exception occurs for subject 8 - no overlap found between the two distributions in the NREM sleep stage. However, the absence of either a complete or partial overlap between the two distributions (PSG-to-In-ear-EEG and PSG-to-PSG similarity scores) does not directly imply a lack of overlapping information between the two different sources, i.e., PSG and in-ear-EEG.

We have to keep in mind that the PSG-to-PSG JSD-FSI similarity scores have been computed mainly to have a reference with which to compare our PSG-to-In-ear-EEG JSD-FSI values. Data from the same source (e.g., PSG signals recorded from the scalp), when compared against each other, should contain close or similar information - resulting in a reference distribution of PSG-to-PSG JSD-FSI similarity scores. Thus, the new in-ear channel, when compared against each of the PSG data sources - i.e., channels derived from the scalp - should result in JSD-FSI values close if not equal to our PSG based reference.

To further analyze and to better interpret the results in Figure 10, Figure 11, and Figure 12 - in Supplementary Analysis - we report, for each sleep stage, the JSD-FSI similarity score reference distributions computed separately for each PSG data source, i.e., scalp-EEG-to-scalp-EEG (Figure S3, Figure S4 and Figure S5) and EOG-to-EOG (Figure S6, Figure S7 and Figure S8).

In Figures S3-S5 we want to investigate whether the in-ear-EEG shows information similar to the one from the scalp-EEG - by comparing the scalp-EEG-to-In-ear-EEG JSD-FSI distributions with scalp-EEG-to-scalp-EEG reference distributions.

In Figures S6-S8 we want to investigate whether the in-ear-EEG shows information similar to the one from the EOG - by comparing the EOG-to-In-ear-EEG JSD-FSI distributions with EOG-to-EOG reference distributions.

As highlighted in the overlapping area in purple, the similarity between the in-ear-EEG and the scalp-EEG channels is higher compared to the one between the in-ear-EEG and the EOG channels - the scalp-EEG-to-In-ear-EEG JSD-FSI distributions overlap with their corresponding reference distributions (Scalp-EEG-to-Scalp-EEG similarity scores) for almost all subjects and in all the sleep stages.

## Discussion

While evaluating the agreement between the PSG and in-ear-EEG hypnograms scored by the same scorer expert (hypnogram-based comparison analysis), we found a high intra-scorer variability. The high variability - or inconsistency between the two different sources - is mainly due to the great uncertainty the scorers had in evaluating the in-ear-EEG signals (see the inter-scorer variability analysis). The Fleiss' kappa values between the in-ear-EEG scorers are on average lower - and not consistent - compared to the ones computed on the PSG scorers. We may infer that the in-ear-EEG recordings are harder to score than traditional PSG signals. However, the heightened scoring complexity may not stem from the substandard quality of the in-ear-EEG signal, rather from the innovative nature of the EEG source captured from our ears - distinctly divergent from what scoring experts are used to look at. The main constraint - compared to the traditional PSG based scoring procedure - is that the scorers are assigning the sleep stages just relying on a single in-ear-EEG channel. The scorers - hence the physicians - are used to score our sleep considering simultaneously information that comes from different channels. When scoring in-ear-EEG they should consider different protocols.

In our feature-based comparison analysis, we showed a substantial similarity in terms of JSD-FSI score - on average 0.79 ± 0.06 in awake, 0.77 ± 0.07 in NREM, and 0.67 ± 0.10 in REM - between the two different sources. The in-ear-EEG signals are retaining information (in time- and frequency- domain) close to the ones we usually extrapolate through standard PSG derivations. This latter claim is in contrast to what we found following our alternative approach, i.e., the hypnogram-based comparison analysis, which mainly relies on the experience and knowledge of the scoring experts. However, considering the agreement metrics between PSG and in-ear EEG consensus, it is evident that the discrepancy between the sleep scoring of these two sources is primarily due to the REM stage.

The robustness of the JSD-FSI similarity scores - and the significance of the similarity values per se - is further validated showing the clear alignment between the PSG-to-In-ear-EEG and PSG-to-PSG JSD-FSI score distributions. The similarity between the in-ear-EEG and any PSG derivation is close to the one we would find between any pair of standard scalp PSG derivations.

To summarize, our results perfectly align with previous works [2, 8-13] emphasizing the use of mobile ear-EEG solutions as promising alternatives to standard PSG, as the comparison between the sleep-scoring performances on ear-EEG and PSG recordings indirectly represents a measure of similarity between these two sources in the context of sleep analysis.

Some of these studies [9, 12, 13] also highlight the difficulties in scoring the REM stage compared to other classes. Such a difficulty is consistent with our analysis, as we observe significant changes in JSD-FSI score between the sleep stages, with significantly greater values for the awake stage if compared to NREM and REM ones, and for the NREM stage with respect to the REM state. Therefore, the outcomes with the smallest similarity scores for REM sleep are in line with what we already knew to date: the in-ear-EEG sensors may be not enough in distinguishing REM sleep stage - additional information from EOG signals is needed [3, 10, 16]. This claim is further supported by the results in Figures S3-S8 - where the similarity of the in-ear-EEG with EOG derivations was, overall, lower than the one with scalp-EEG derivations.

The main limitation of this preliminary study is that we cannot make any comprehensive consideration regarding JSD-FSI consistency between subjects or channels - i.e., spatial distribution. There is a need to further validate the proposed methodology on a higher number of recordings - eventually involving subjects affected by different sleep disorders - increasing data heterogeneity.

## Funding

The present work has been partially funded by IDUN GESSE (Guardian Earbuds Sleep Scorer Expert), INNO-ICT 65141.1, from the Innovation Cheque funding program.

## Conflict of Interest

SUPSI authors are responsible for the research, and they conducted the analysis independently. IDUN Company representative contribution was in offering detailed information about the device and the dataset collected in a previous study.

## Data availability

The dataset from IDUN (BASEC Nr. Req-2022-00105) is not publicly available. The individual identities of participants cannot be inferred from the in-ear-EEG and/or PSG signals. All the data used in this study were collected in accordance with ethical guidelines and with informed consent from the participants. The data are available on request for non-commercial purposes (legal conditions ensuring data privacy and official ethical guidelines compliance will be defined in a "data transfer agreement document", together with a description of the analysis project).

## Code availability

To ensure the reproducibility of our analysis, the code has been made openly accessible and is hosted on [GitHub](GitHub).

## Author contributions

All the authors contributed to the design of the study; G.P. and L.F. contributed equally to the study; G.P. under L.F. guidance implemented the system and conducted all the experiments; L.F., G.P., and F.D.F. wrote the paper with feedback from G.M., M.B., M.M., A.T. and V.A.; M.W. designed the quality assurance study, managed participant screening, data collection, and performed initial data analysis to assess data quality; E.M. designed the PSG and in-ear EEG hardware setup with study participants, collected and scored the sleep recordings, and organized/instructed the two additional sleep scorers; F.P.B.R. assisted in setting up the sleep lab and study site, participant recruitment, and data collection; all authors approved the final paper.

# References

[1] Stochholm A, Mikkelsen K, and Kidmose P. Automatic sleep stage classification using ear-EEG. 2016 38th Annual International Conference of the IEEE Engineering in Medicine and Biology Society (EMBC). IEEE. 2016; 4751–4. doi: 10.1109/EMBC.2016.7591789.

[2] Tabar YR, Mikkelsen KB, Shenton N, Kappel SL, Bertelsen AR, Nikbakht R, Toft HO, Henriksen CH, Hemmsen MC, Rank ML, et al. At-home sleep monitoring using generic ear-EEG. Frontiers in Neuroscience 2023; 17:987578. doi: 10.3389/fnins.2023.987578.

[3] Zibrandtsen I, Kidmose P, Otto M, Ibsen J, and Kjaer T. Case comparison of sleep features from ear-EEG and scalp-EEG. Sleep Science 2016; 9:69–72. doi: 10.1016/j.slsci.2016.05. 006.

[4] Berry RB, Brooks R, Gamaldo CE, Harding SM, Marcus C, Vaughn BV, et al. The AASM manual for the scoring of sleep and associated events. Rules, Terminology and Technical Specifications, Darien, Illinois, American Academy of Sleep Medicine 2012; 176:2012.

[5] Fiorillo L, Puiatti A, Papandrea M, Ratti PL, Favaro P, Roth C, Bargiotas P, Bassetti CL, and Faraci FD. Automated sleep scoring: A review of the latest approaches. Sleep Medicine Reviews 2019; 48:101204. doi: 10.1016/j.smrv.2019.07.007.

[6] Imtiaz SA. A systematic review of sensing technologies for wearable sleep staging. Sensors 2021; 21:1562. doi: 10.3390/s21051562.

[7] Looney D, Kidmose P, Park C, Ungstrup M, Rank ML, Rosenkranz K, and Mandic DP. The in-the-ear recording concept: User-centered and wearable brain monitoring. IEEE Pulse 2012; 3:32–42. doi: 10.1109/MPUL.2012. 2216717.

[8] Nakamura T, Goverdovsky V, Morrell MJ, and Mandic DP. Automatic sleep monitoring using ear-EEG. IEEE journal of translational engineering in health and medicine 2017; 5:1–8. doi: 10.1109/JTEHM.2017.2702558.

[9] Mikkelsen KB, Villadsen DB, Otto M, and Kidmose P. Automatic sleep staging using ear-EEG. Biomedical engineering online 2017; 16:1–15. doi: 10.1186/s12938-017-0400-5.

[10] Mikkelsen KB, Ebajemito JK, Bonmati-Carrion MA, Santhi N, Revell VL, Atzori G, Della Monica C, Debener S, Dijk DJ, Sterr A, et al. Machine learning-derived sleep-wake staging from around-the-ear electroencephalogram outperforms manual scoring and actigraphy. Journal of Sleep Research 2019; 28:e12786. doi: 10.1111/jsr.12786.

[11] Mikkelsen KB, Tabar YR, Kappel SL, Christensen CB, Toft HO, Hemmsen MC, Rank ML, Otto M, and Kidmose P. Accurate whole-night sleep monitoring with dry-contact ear-EEG. Scientific Reports 2019; 9:16824. doi: 10.1038/s41598-019-53115-3.

[12] Nakamura T, Alqurashi YD, Morrell MJ, and Mandic DP. Hearables: automatic overnight sleep monitoring with standardized in-ear-EEG sensor. IEEE Transactions on Biomedical Engineering 2019; 67:203–12. doi: 10.1109/TBME.2019.2911423.

[13] Tabar YR, Mikkelsen KB, Rank ML, Hemmsen MC, Otto M, and Kidmose P. Ear-EEG for sleep assessment: a comparison with actigraphy and PSG. Sleep and Breathing 2021; 25:1693–705. doi: 10.1007/s11325-020-02248-1.

[14] Jørgensen SD, Zibrandtsen IC, and Kjaer TW. Ear-EEG-based sleep scoring in epilepsy: A comparison with scalp-EEG. Journal of Sleep Research 2020; 29:e12921. doi: 10.1111/jsr.12921.

[15] Kjaer TW, Rank ML, Hemmsen MC, Kidmose P, and Mikkelsen K. Repeated automatic sleep scoring based on ear-EEG is a valuable alternative to manually scored polysomnography. PLOS Digital Health 2022; 1:e0000134. doi: 10.1371/journal.pdig.0000134.

[16] Silva Souto CF da, Pätzold W, Wolf KI, Paul M, Matthiesen I, Bleichner MG, and Debener S. Flex-printed ear-EEG sensors for adequate sleep staging at home. Frontiers in Digital Health 2021; 3:688122. doi: 10.3389/fdgth.2021.688122.

[17] Buysse DJ, Reynolds III CF, Monk TH, Berman SR, and Kupfer DJ. The Pittsburgh Sleep Quality Index: a new instrument for psychiatric practice and research. Psychiatry Research 1989; 28:193–213. doi: 10.1016/0165-1781(89)90047-4.

[18] Åkerstedt T and Gillberg M. Subjective and objective sleepiness in the active individual. International journal of neuroscience 1990; 52:29–37. doi: 10.3109/00207459008994241.

26[19] Fiorillo L, Pedroncelli D, Agostini V, Favaro P, and Faraci FD. Multi-scored sleep databases: how to exploit the multiple-labels in automated sleep scoring. Sleep 2023; 46:zsad028. doi: 10.1093/sleep/zsad028.

[20] Guillot A, Sauvet F, During EH, and Thorey V. Dreem open datasets: Multiscored sleep datasets to compare human and automated sleep staging. IEEE transactions on neural systems and rehabilitation engineering 2020; 28:1955– 65. doi: 10.1109/TNSRE.2020.3011181.

[21] Lee YJ, Lee JY, Cho JH, and Choi JH. Interrater reliability of sleep stage scoring: a meta-analysis. Journal of Clinical Sleep Medicine 2022; 18:193–202. doi: 10.5664/jcsm.9538.

[22] Delgado-Bonal A and Marshak A. Approximate entropy and sample entropy: A comprehensive tutorial. Entropy 2019; 21:541. doi: 10.3390/ e21060541.

[23] Min J, Wang P, and Hu J. Driver fatigue detection through multiple entropy fusion analysis in an EEG-based system. PLoS one 2017; 12:e0188756. doi: 10.1371/journal.pone.0188756.

[24] Molina-Picó A, Cuesta-Frau D, Aboy M, Crespo C, Miró-Martınez P, and Oltra-Crespo S. Comparative study of approximate entropy and sample entropy robustness to spikes. Artificial intelligence in medicine 2011; 53:97– 106. doi: 10.1016/j.artmed.2011.06.007.

[25] Wu J, Zhou T, and Li T. Detecting epileptic seizures in EEG signals with complementary ensemble empirical mode decomposition and extreme gradient boosting. Entropy 2020; 22:140. doi: 10.3390/e22020140.

[26] Krishnan PT, Raj ANJ, Balasubramanian P, and Chen Y. Schizophrenia detection using MultivariateEmpirical Mode Decomposition and entropy measures from multichannel EEG signal. Biocybernetics and Biomedical Engineering 2020; 40:1124–39. doi: 10.1016/j.bbe.2020.05.008.

[27] Deng B, Cai L, Li S, Wang R, Yu H, Chen Y, and Wang J. Multivariate multi-scale weighted permutation entropy analysis of EEG complexity for Alzheimer's disease. Cognitive neurodynamics 2017; 11:217–31. doi: 10.1007/s11571-016-9418-9.

[28] Aboy M, Hornero R, Abásolo D, and Álvarez D. Interpretation of the LempelZiv complexity measure in the context of biomedical signal analysis. IEEE transactions on biomedical engineering 2006; 53:2282–8. doi: 10.1109/TBME.2006.883696.

[29] Lee JM, Kim DJ, Kim IY, Park KS, and Kim SI. Detrended fluctuation analysis of EEG in sleep apnea using MIT/BIH polysomnography data. Computers in biology and medicine 2002; 32:37–47. doi: 10.1016/s0010-4825(01)00031-2.

[30] Vallat R and Walker MP. An open-source, high-performance tool for automated sleep staging. Elife 2021; 10:e70092. doi: 10.7554/eLife.70092.

[31] Hjorth B. EEG analysis based on time domain properties. Electroencephalography and clinical neurophysiology 1970; 29:306–10. doi: 10.1016/0013- 4694(70)90143-4.

[32] Lal U, Mathavu Vasanthsena S, and Hoblidar A. Temporal Feature Extraction and Machine Learning for Classification of Sleep Stages Using Telemetry Polysomnography. Brain Sciences 2023; 13:1201. doi: 10.3390/brainsci13081201.

[33] Yan R, Zhang C, Spruyt K, Wei L, Wang Z, Tian L, Li X, Ristaniemi T, Zhang J, and Cong F. Multi-modality of polysomnography signals' fusion for automatic sleep scoring. Biomedical Signal Processing and Control 2019; 49:14–23. doi: 10.1016/j.bspc.2018.10.001.

[34] Hadjidimitriou S, Zacharakis A, Doulgeris P, Panoulas K, Hadjileontiadis L, and Panas S. Sensorimotor cortical response during motion reflecting audiovisual stimulation: evidence from fractal EEG analysis. Medical & biological engineering & computing 2010; 48:561–72. doi: 10.1007/s11517-010-0606-1.

[35] Asirvadam VS, Yusoff MZ, et al. Fractal dimension and power spectrum of electroencephalography signals of sleep inertia state. IEEE Access 2019; 7:185879–92. doi: 10.1109/ACCESS.2019.2960852.

[36] Welch P. The use of fast Fourier transform for the estimation of power spectra: a method based on time averaging over short, modified periodograms. IEEE Transactions on audio and electroacoustics 1967; 15:70–3. doi: 10.1109/TAU.1967.1161901.

[37] Geng D, Wang C, Fu Z, Zhang Y, Yang K, and An H. Sleep EEG-Based Approach to Detect Mild Cognitive Impairment. Frontiers in Aging Neuroscience 2022; 14:865558. doi: 10.3389/fnagi.2022. 865558.

[38] Huang D, Lin P, Fei DY, Chen X, and Bai O. Decoding human motor activity from EEG single trials for a discrete two-dimensional cursor control. Journal of Neural Engineering 2009; 6:046005. doi: 10.1088/1741-2560/6/4/046005.

[39] Memar P and Faradji F. A novel multi-class EEG-based sleep stage classification system. IEEE Transactions on Neural Systems and Rehabilitation Engineering 2017; 26:84–95. doi: 10.1109/TNSRE.2017.2776149.

[40] Mera-Gaona M, López DM, and Vargas-Canas R. An Ensemble feature selection approach to identify relevant features from EEG signals. Applied Sciences 2021; 11:6983. doi: 10.3390/app11156983.

[41] Hassan AR, Bashar SK, and Bhuiyan MIH. Automatic classification of sleep stages from single-channel electroencephalogram. 2015 annual IEEE India conference (INDICON). IEEE. 2015; 1–6. doi: 10.1109/INDICON.2015.7443756.

[42] Sharma N, Kolekar M, Jha K, and Kumar Y. EEG and cognitive biomarkers based mild cognitive impairment diagnosis. Irbm 2019; 40:113–21. doi: 10.1016/j.irbm.2018.11.007.

[43] Choudhury AR, Ghosh A, Pandey R, and Barman S. Emotion recognition from speech signals using excitation source and spectral features. 2018 IEEE Applied Signal Processing Conference (ASPCON). IEEE. 2018; 257–61. doi: 10.1109/ASPCON.2018.8748626.

[44] Mitra P, Murthy C, and Pal SK. Unsupervised feature selection using feature similarity. IEEE transactions on pattern analysis and machine intelligence 2002; 24:301–12. doi: 10.1109/34.990133.

[45] Solorio-Fernández S, Carrasco-Ochoa JA, and Martınez-Trinidad JF. A review of unsupervised feature selection methods. Artificial Intelligence Review 2020; 53:907–48. doi: 10.1007/s10462-019-09682-y.

[46] Yan X, Nazmi S, Erol BA, Homaifar A, Gebru B, and Tunstel E. An efficient unsupervised feature selection procedure through feature clustering. Pattern Recognition Letters 2020; 131:277–84. doi: 10.1016/j.patrec.2019.12.022.

[47] Zhao Z, Wang L, Liu H, and Ye J. On similarity preserving feature selection. IEEE Transactions on Knowledge and Data Engineering 2011; 25:619–32. doi: 10.1109/TKDE.2011.222.

[48] Klumpe H, Lugagne JB, Khalil AS, and Dunlop M. Deep neural networks for predicting single cell responses and probability landscapes. bioRxiv 2023; 2023–6. doi: 10.1021/acssynbio.3c00203.

[49] Dragalin V, Fedorov V, Patterson S, and Jones B. Kullback–Leibler divergence for evaluating bioequivalence. Statistics in medicine 2003; 22:913–30. doi: 10.1002/sim.1451.

[50] Bullmann M, Fetzer T, Ebner F, Deinzer F, and Grzegorzek M. Fast kernel density estimation using Gaussian filter approximation. 2018 21st International Conference on Information Fusion (FUSION). IEEE. 2018; 1233–40. doi: 10.23919/ICIF.2018.8455686.

# Tables

**Table 1**

Number and percentage of 30-second epochs per sleep stage (i.e., the result of the consensus reached by the three different scorers) for both PSG and in-ear-EEG based scoring procedure, and the intersection (∩) computed on the labels coming from the two different sources.

|  | *W* | *NREM* | *REM* | *Total* |
|---|---|---|---|---|
| **PSG** | 344 (7.5%) | 3469 (75.9%) | 755 (16.5%) | 4568 |
| **In-ear-EEG** | 277 (6.1%) | 3768 (82.5%) | 523 (11.4%) | 4568 |
| **PSG ∩ In-ear-EEG** | 236 (6.0%) | 3308 (84.0%) | 392 (10.0%) | 3936 |

**Table 2**

Average precision, recall, and F1-score metrics evaluated across all subjects between PSG and in-ear-EEG consensus, considering the former as the gold-standard.

|  | *W* | *NREM* | *REM* |
|---|---|---|---|
| **Precision** | 0.84 ± 0.16 | 0.88 ± 0.05 | 0.65 ± 0.35 |
| **Recall** | 0.80 ± 0.21 | 0.95 ± 0.03 | 0.47 ± 0.27 |
| **F1-score** | 0.79 ± 0.14 | 0.91 ± 0.03 | 0.53 ± 0.28 |

# Figure Captions

**Figure 1.** Schematic layout of the quality assurance study and the data collection procedure.

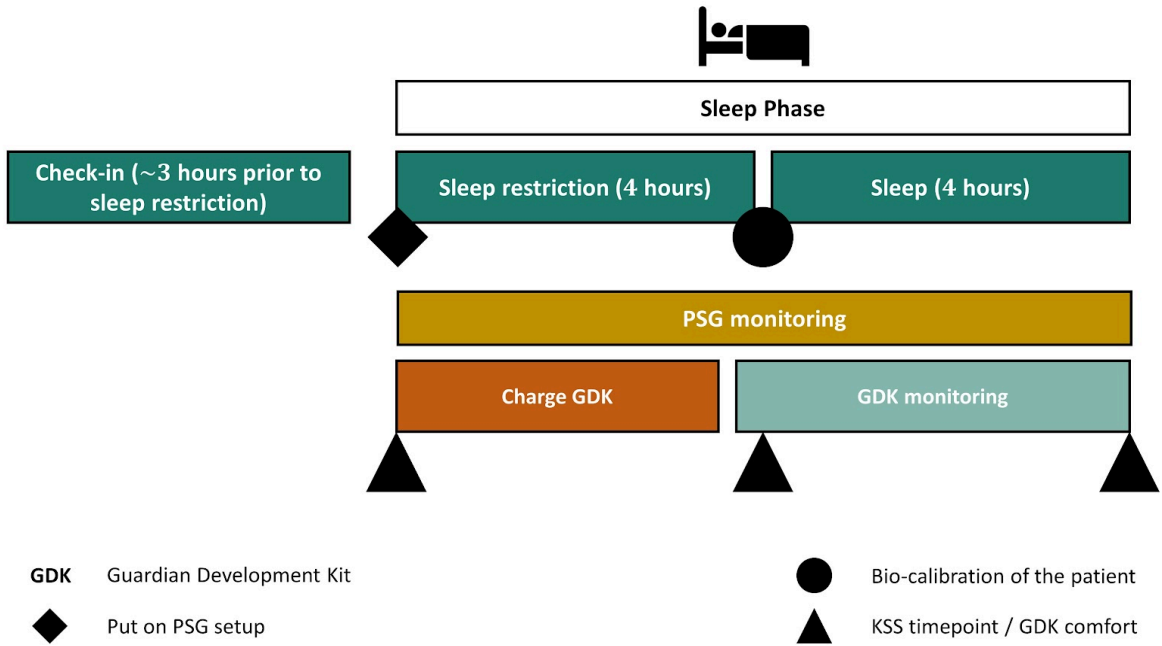

**Figure 2.** Devices employed in the data collection: a) SOMNOmedics SOMNOscreen plus system for PSG data with the EXG configuration, i.e., including six scalp electrodes, EOG, and ECG signal monitoring; b) Guardian Development Kit (GDK) hardware including ear tips, earpieces, and brain box used to record in-ear-EEG data.

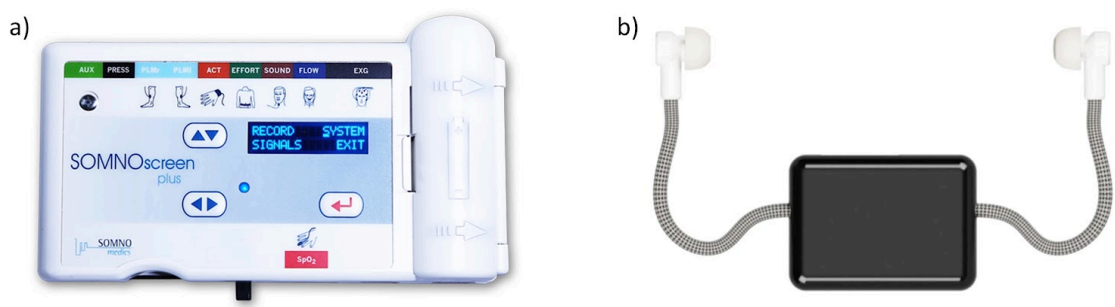

**Figure 3.** Workflow for evaluating the similarity between the signals recorded from two different channels, including feature extraction and feature selection, separately for each sleep stage; and the comparison between feature distributions using the Jensen–Shannon divergence before the assessment of the similarity-scores, individually for each sleep stage and for each subject. In detail, a) refers to the comparison between one in-ear-EEG and one PSG channel (either scalp-EEG or EOG channels); while b) illustrates the analysis between two PSG channels (either scalp-EEG or EOG channels). An example of similarity-scores distribution for awake, NREM, and REM classes is included for both case studies.

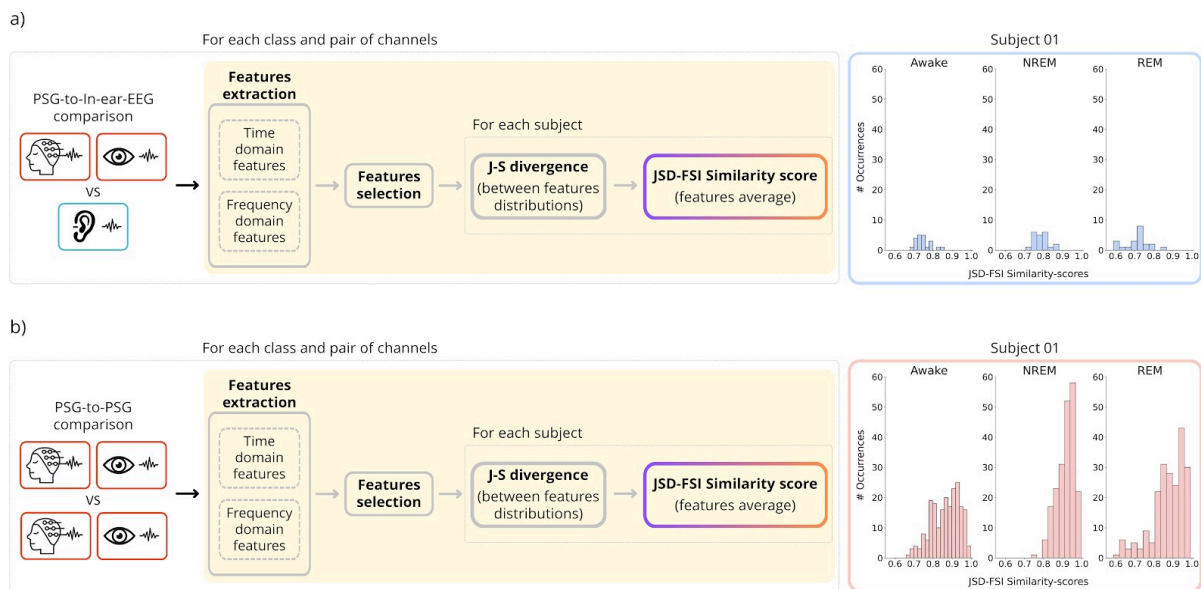

**Figure 4.** Intra-scorer variability (multi-source-scored dataset). Boxplot distribution of the Cohen's kappa values computed for each recording/subject between the PSG and in-ear-EEG hypnograms - for each scorer.

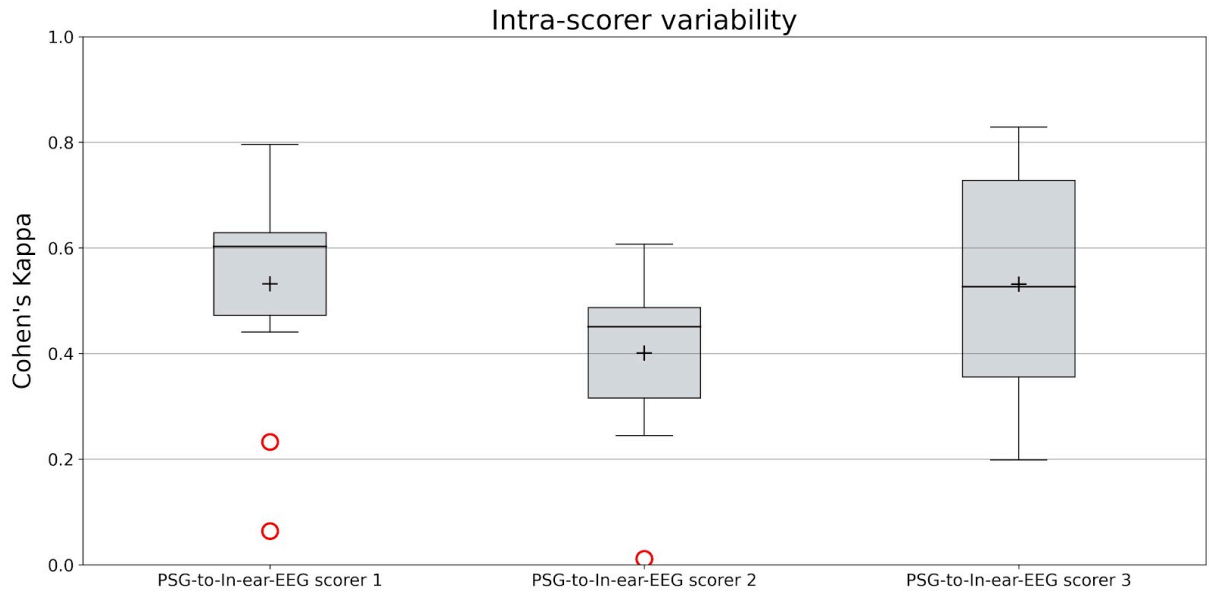

**Figure 5.** Inter-scorer variability (multi-source-scored dataset). Boxplot distributions of the Fleiss' kappa values computed for each recording/subject between the three scorer experts for in-ear-EEG (in blue) and PSG (in red) signals.

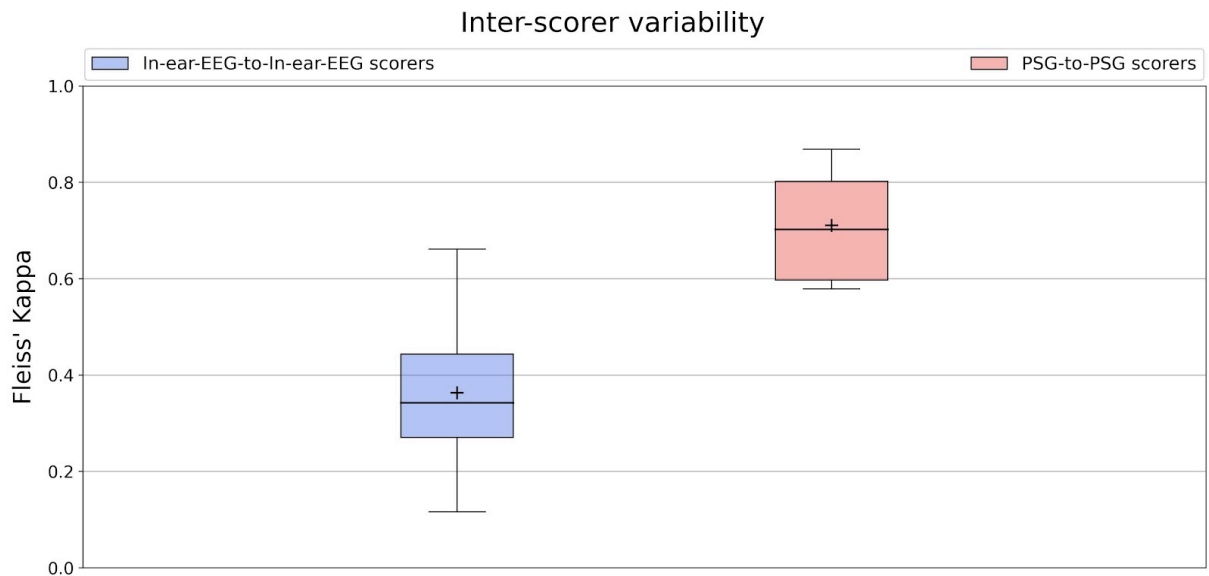

**Figure 6.** Color-coded heatmap showing the selection frequency of each extracted feature across the various subsets {$D^q$, $D^{CH1}$}, separately for each sleep stage, reporting warmer colors for higher frequencies.

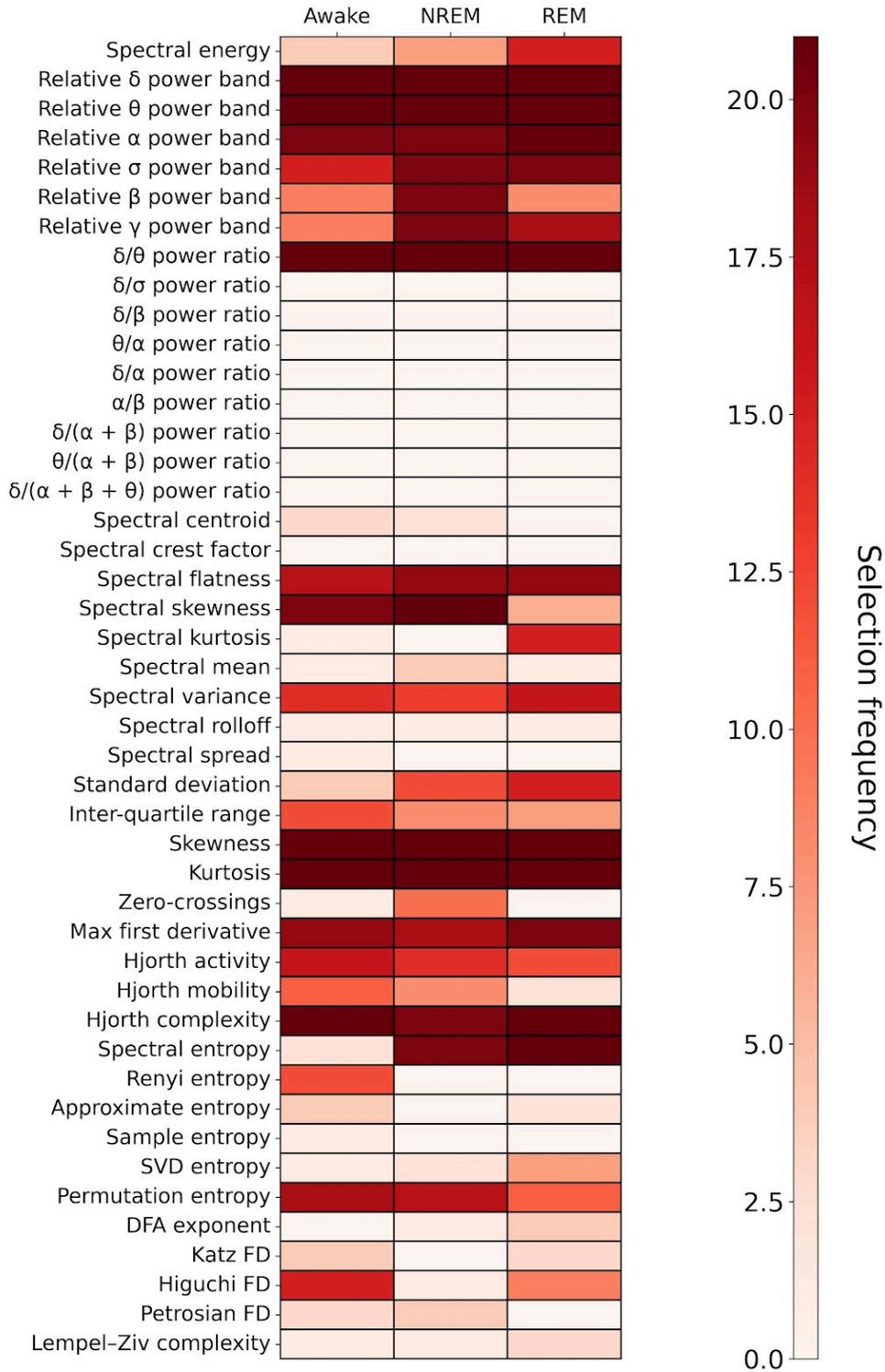

**Figure 7.** Head topography plots of JSD-FSI similarity-scores of the in-ear-EEG with respect to the PSG channels in the $Q$ set (EEG and EOG channels) - for each subject in the awake stage. Unipolar channels are represented as markers; bipolar channels are represented as lines. The mean and the standard deviation of the similarity-scores distribution computed for each subject are also included. No JSD-FSI similarity-scores are reported for the channel M2 of subjects 3 and 6.

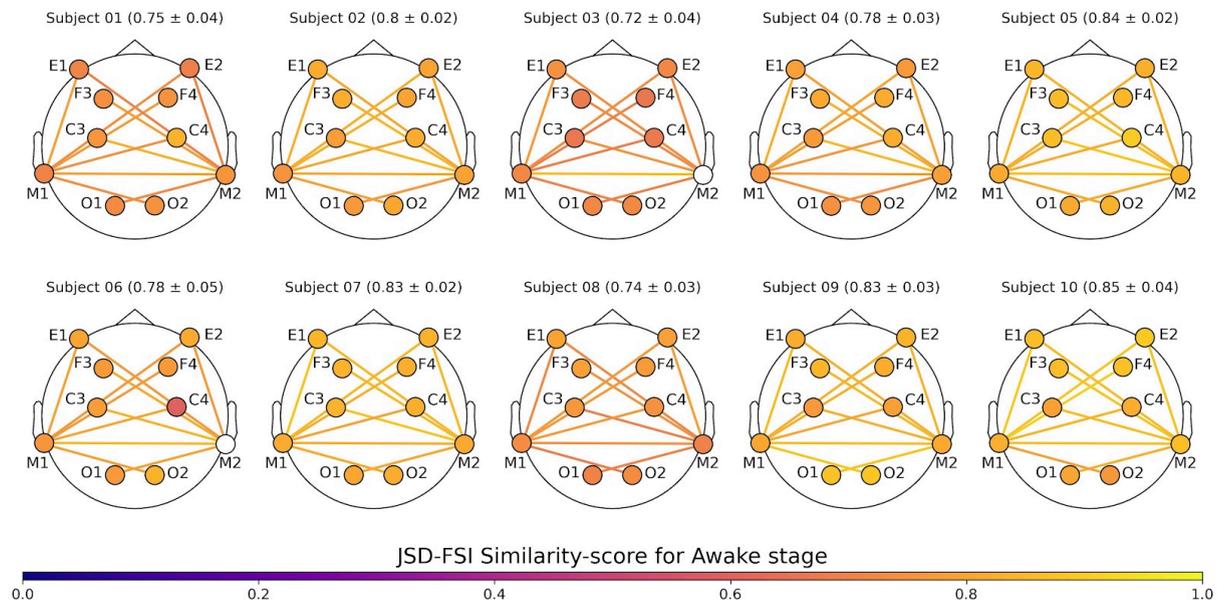

**Figure 8.** Head topography plots of JSD-FSI similarity-scores of the in-ear-EEG with respect to the PSG channels in the $Q$ set (EEG and EOG channels) - for each subject in the NREM stage. Unipolar channels are represented as markers; bipolar channels are represented as lines. The mean and the standard deviation of the similarity-scores distribution computed for each subject are also included. No JSD-FSI similarity-scores are reported for the channel M2 of subjects 3 and 6.

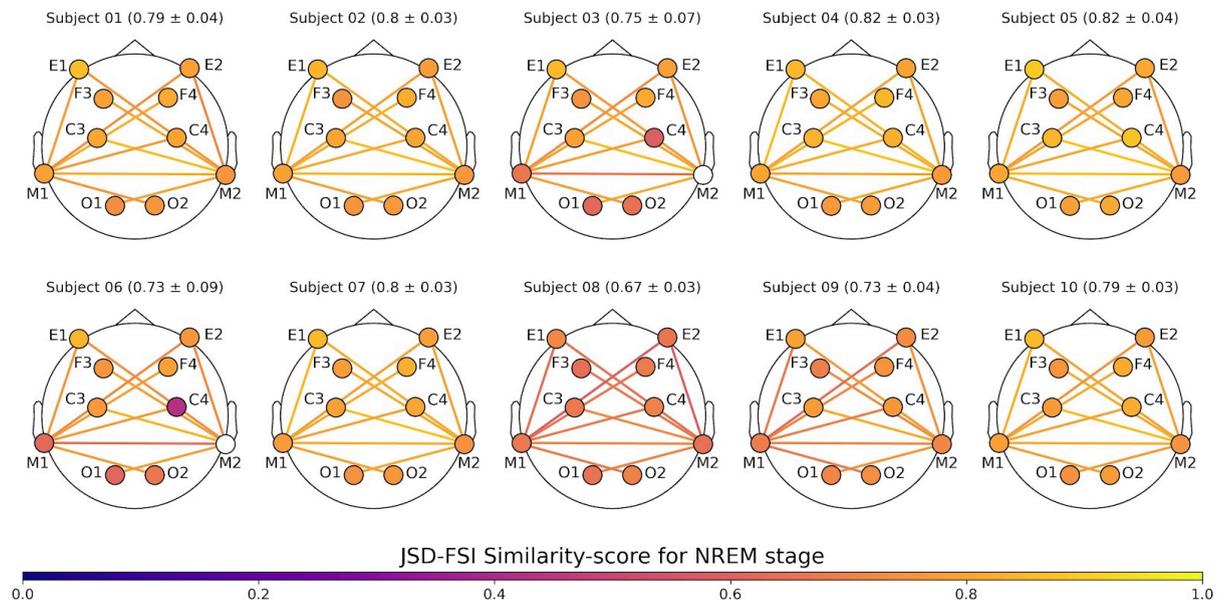

**Figure 9.** Head topography plots of JSD-FSI similarity-scores of the in-ear-EEG with respect to the PSG channels in the $Q$ set (EEG and EOG channels) - for each subject in the REM sleep stage. Unipolar channels are represented as markers; bipolar channels are represented as lines. The mean and the standard deviation of the similarity-scores distribution computed for each subject are also included. No JSD-FSI similarity-scores are reported for subjects 3 and 8 nor for the channel M2 of subject 6.

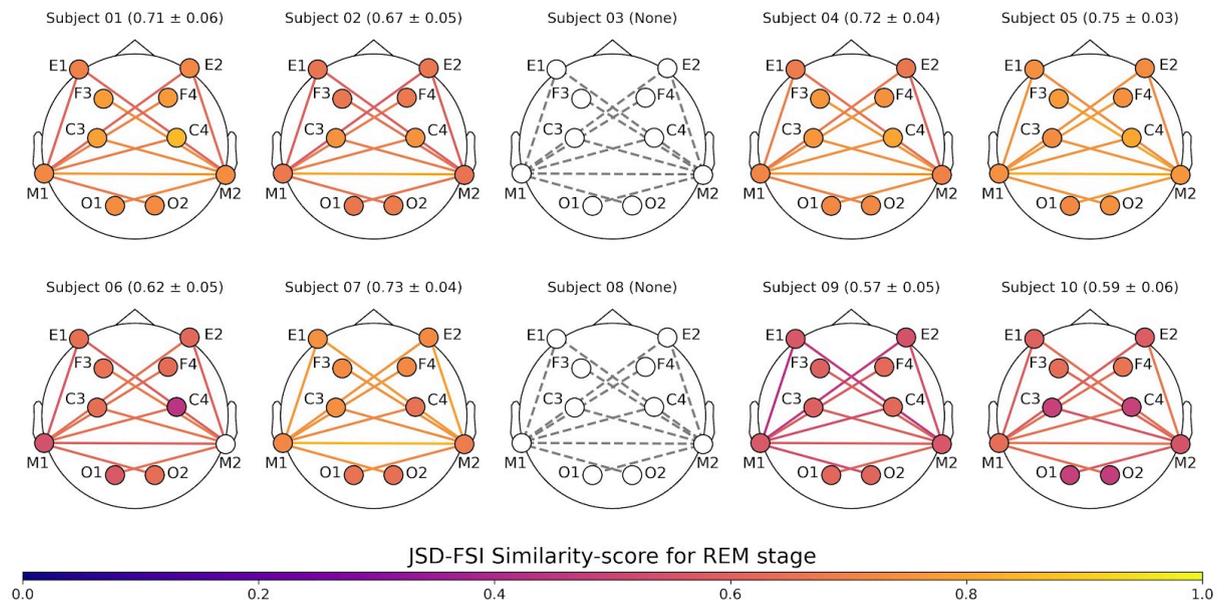

**Figure 10.** JSD-FSI similarity-scores distributions, i.e., distributions derived from the PSG-to-In-ear-EEG (histograms in blue) and PSG-to-PSG (histograms in red) comparisons - for each subject in the awake stage.

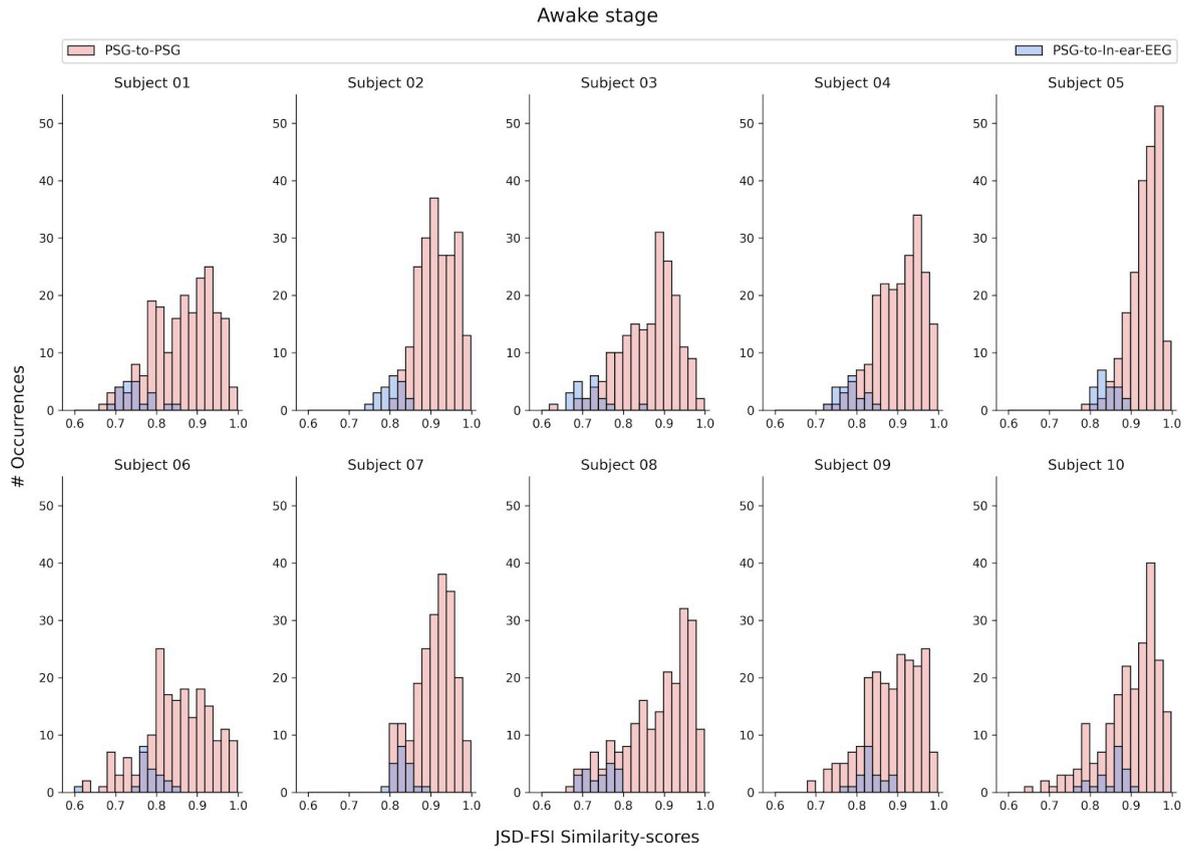

**Figure 11.** JSD-FSI similarity-scores distributions, i.e., distributions derived from the PSG-to-In-ear-EEG (histograms in blue) and PSG-to-PSG (histograms in red) comparisons - for each subject in the NREM sleep stage.

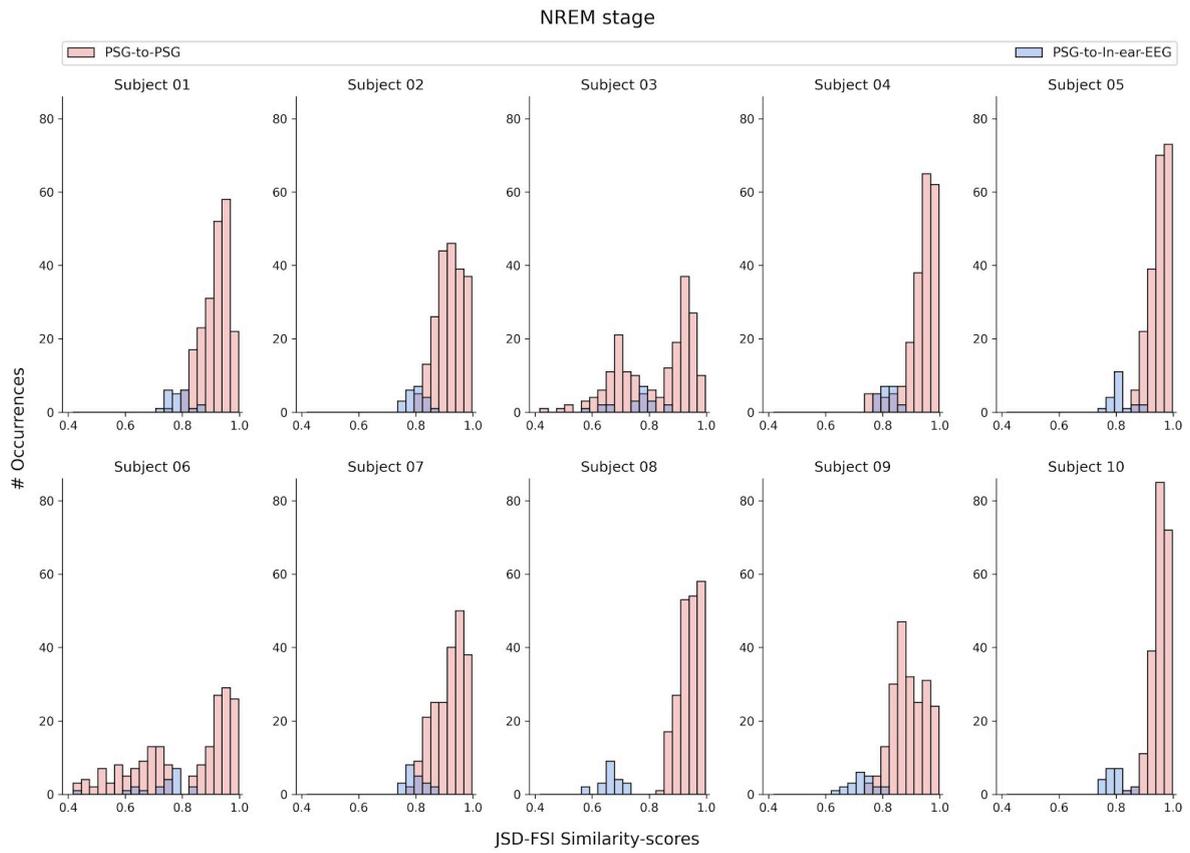

**Figure 12.** JSD-FSI similarity-scores distributions, i.e., distributions derived from the PSG-to-In-ear-EEG (histograms in blue) and PSG-to-PSG (histograms in red) comparisons - for each subject in the REM sleep stage. No JSD-FSI similarity-scores are reported for subjects 3 and 8.

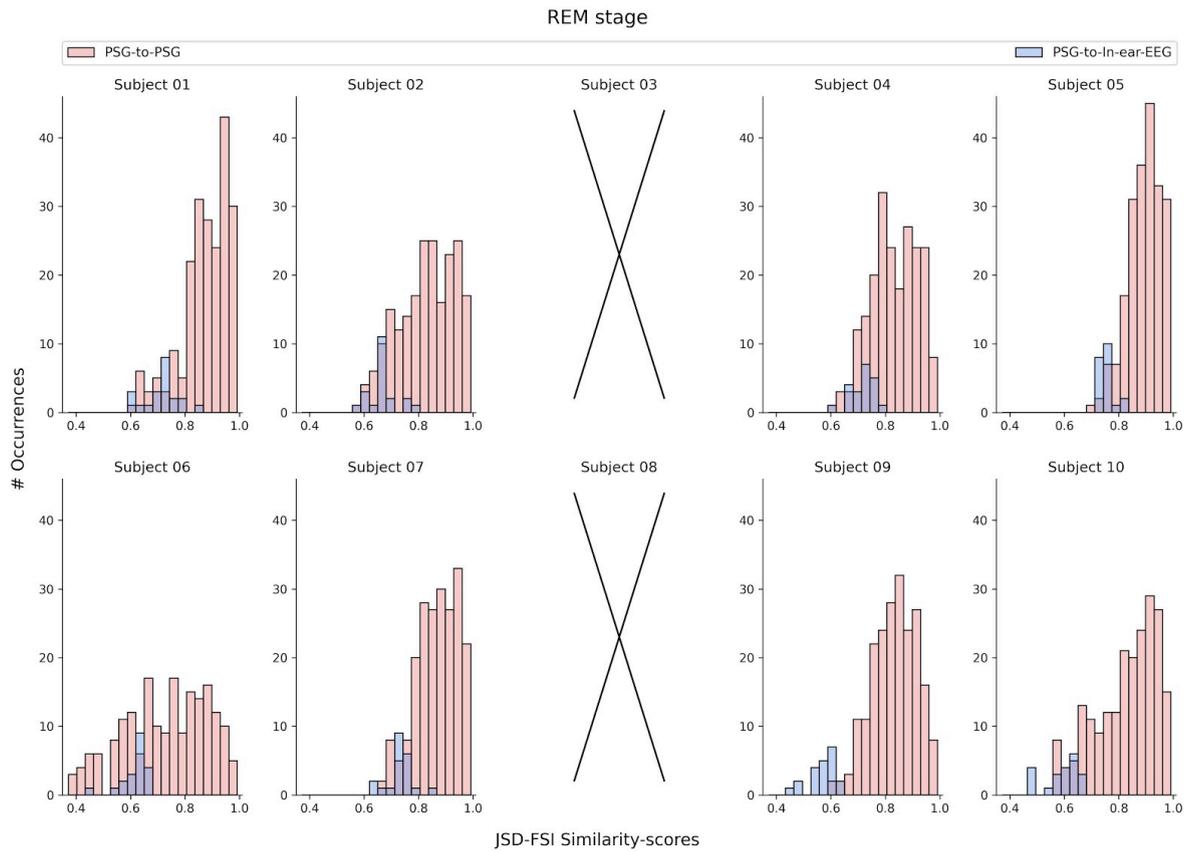

# SUPPLEMENTARY MATERIAL

# Comparison analysis between standard polysomnographic data and in-ear-EEG signals: A preliminary study

Gianpaolo Palo[1,2,†], Luigi Fiorillo[1,*,†], Giuliana Monachino[1,3], Michal Bechny[1,3], Michel Wälti[4], Elias Meier[4], Francesca Pentimalli Biscaretti di Ruffia[4], Mark Melnykowycz[4], Athina Tzovara[3], Valentina Agostini[2], and Francesca Dalia Faraci[1]

[1]Institute of Digital Technologies for Personalized Healthcare (MeDiTech), Department of Innovative Technologies, University of Applied Sciences and Arts of Southern Switzerland, Lugano, Switzerland; [2]Department of Electronics and Telecommunications, Politecnico di Torino, Torino, Italy; [3]Institute of Computer Science, University of Bern, Bern, Switzerland; [4]IDUN Technologies AG, Glattpark, Switzerland.

Institution where work was performed: Institute of Digital Technologies for Personalized Healthcare (MeDiTech), Department of Innovative Technologies, University of Applied Sciences and Arts of Southern Switzerland, Lugano, Switzerland.

†These authors contributed equally to this work.

*Corresponding author - Email: luigi.fiorillo@supsi.ch.

# SUPPLEMENTARY ANALYSES

## Time-domain features

In the list below we report mathematical details on how to compute each feature.

- ***Standard deviation*** and ***interquartile range*** are statistical measures of dispersion, and are exploited in our study to describe the variation of the electrical activity in the brain.

- ***Skewness*** and ***kurtosis*** are the third and fourth central moments in statistics, and are used to characterize the shape of the EEG signals in terms of asymmetry relative to the mean and heaviness of the tails compared to a normal distribution.

- The ***maximum first derivative*** and the ***number of zero-crossings*** give information about the depolarization of an EEG recording. The former outlines the highest rate of depolarization in the signal, while the latter estimates its frequency of sign-changes.

- ***Approximate entropy*** and ***sample entropy*** are two time series regularity metrics. In general, entropy measures evaluate the complexity of a time series by estimating its intrinsic disorder. The higher the entropy of a signal, the less predictable its evolution. The approximate entropy, *ApEn* (1), quantifies the logarithmic likelihood that a signal of length $N$ repeats itself always the same within a certain tolerance $r$. Its estimation includes dividing the signal in sub-segments first of length $d$ and then $(d + 1)$ before evaluating the correlation integral, $C(r)$, which expresses how many times each *i-th* sub-segment is akin to all the others [22, 24].

$$ApEn(d, r, N) = \frac{\sum_{i=1}^{N-d+1} \log\left[C_i^d(r)\right]}{N-d+1} - \frac{\sum_{i=1}^{N-(d+1)+1} \log\left[C_i^{d+1}(r)\right]}{N-(d+1)+1} \qquad (1)$$

The sample entropy, *SpEn* (2), can be taken as a simpler version of the approximate entropy. The difference lies in no longer considering the self-matching for each sub-segment [22].

$$SpEn(d, r, N) = -\frac{1}{N-d+1} \log\left[\frac{\sum_{i=1}^{N-d+1} C_i^{d+1}(r)}{\sum_{i=1}^{N-d+1} C_i^d(r)}\right] \qquad (2)$$

Parameters $d$ and $r$ are respectively set to 2 and $0.2 \cdot SD$ (*SD*: standard deviation of the signal) for both *ApEn* and *SpEn* [23-25].

- **Singular Value Decomposition entropy (SVDEn)**, quantifies the complexity of the time series based on the number of eigenvectors needed for its adequate representation. Mathematically, starting from a signal of length $N$, i.e., $[x_1, x_2, ..., x_N]$ some delay vectors $y(i) = \left[x_i, x_{i+\tau}, ..., x_{i+(d_E-1)\tau}\right]$ are built, having $i = 1, 2, ..., N$. Two parameters are involved, i.e., the embedded dimension ($d_E$) and the delay ($\tau$), which are respectively set to 3 and 1, as recommended for short duration EEG signals [25, 26]. SVD factorization is then performed on the embedded matrix $Y = \left[y(1), y(2), ..., y(N-(d_E-1)\tau)\right]$, resulting in $M$ singular values $(\sigma_1, \sigma_2, ..., \sigma_M)$. Hence, SVD entropy is computed on normalized singular values, such that $\overline{\sigma}_i = \sigma_i / \sum_{j=1}^{M} \sigma_j$, following (3) [26].

$$SVDEn = -\sum_{i=1}^{M} \overline{\sigma_i} \log_2\left(\overline{\sigma_i}\right) \qquad (3)$$

- **Permutation entropy (PermEn)**, involves the partition of the time series of length $N$ into a matrix of overlapping column vectors of length $D$. The overlap is defined by a delay parameter ($\tau$). In this work, $D = 3$ and $\tau = 1$ [25, 27]. Column vectors are then mapped into permutations ($\pi$) capturing the ordinal rankings of data. Hence, permutation entropy is calculated as shown in (4), where $p_i$ is the relative frequency of the *i-th* permutation ($\pi_i$) and is evaluated as the number of times $\pi_i$ is found in the signal over the total number of sequences [25, 27].

$$SVDEn = -\sum_{i=1}^{D!} p_i \log_2(p_i) \qquad (4)$$

- **Lempel-Ziv complexity (C)**, evaluates the randomness of the time series. First, the signal of length $N$ is transformed into a binary sequence using coarse-graining. This is performed by assigning 0 and 1 to values respectively below and above a threshold. The latter is defined as the median of the window time series analyzed for its robustness to possible outliers [28]. While inspecting the signal from left to right, the overall number of distinct subsequences, $c(N)$, of consecutive characters is assessed. To get a measure that is independent of the sequence length, Lempel-Ziv complexity is defined as the normalized version of $c(N)$ (5).

$$C(N) = c(N) \frac{\log_2(N)}{N} \qquad (5)$$

- **Detrended Fluctuation Analysis (DFA) exponent**, α, measures the degree of long-term statistical dependencies or intrinsic self-similarity in time series. This feature is used to quantify the long-range correlation properties of the EEG signal by evaluating possible self-similar patterns in the electrical activity of the brain. First, the time series $x$ of length $N$ is integrated to define the '*accumulated walk*', $y(k)$ (6).

$$y(k) = \sum_{i=1}^{k} [x(i) - \bar{x}] \qquad (6)$$

where $x(i)$ is the sequence at the *i-th* sample, and $\bar{x}$ is the mean of the entire time series. Hence, $y(k)$ is divided into sub-sequences of equal length $n$, and a least-squares line fitting the data within each sub-sequence is evaluated. The y-coordinate of the latter, $y_n(i)$, describes the local trend of each sub-sequence. The Root-Mean-Square (RMS) fluctuation, $F(n)$ (7), is then calculated by integrating and detrending $y(k)$.

$$F(n) = \sqrt{\frac{1}{N} \sum_{i=1}^{N} \left[y(i) - y_n(i)\right]^2} \qquad (7)$$

The whole process is iterated over all the possible lengths, $n$. The DFA exponent is defined as the slope of the best-fitting line to the distribution of the RMS fluctuation as a function of the segment size on a log-log scale.
Whether $\alpha > 0.5$, the time series is characterized by long-range correlations, while $\alpha < 0.5$ outlines that the signal is *anti-persistent, i.e.,* it shows a negative correlation. The case $\alpha = 0.5$ indicates that changes in time series are random and uncorrelated with each other, thus the signal can be modeled as *white noise* [29, 30].

- **Hjorth parameters**, i.e., **activity** $H_A$, **mobility** $H_M$, **and complexity** $H_C$, are statistical functions based on the first and second derivatives of time series useful to characterize the dynamics of the brain function. Let $x$ be an EEG signal of length $N$ and $\bar{x}$ be its mean value, $H_A$ (8) provides information about its energy; $H_M$ (8) is introduced to assess the variability of its frequency over time; and $H_C$ (8) gives insights into the complexity of its waveform in terms of amplitude and frequency with respect to a common sine wave [31].

$$H_A = var[x(t)] = \frac{\sum_{i=1}^{N}(x_i - \bar{x})^2}{N}; \quad H_M = \sqrt{\frac{var\left[\frac{d}{dt}x(t)\right]}{H_A}}; \quad H_C = \frac{H_M\left[\frac{d}{dt}x(t)\right]}{H_M[x(t)]} \quad (8)$$

- **Katz**, **Higuchi**, and **Petrosian** fractal dimensions estimate the fractality of the time series by measuring its complexity and self-similarity over different scales of observation [34].

According to Katz fractal dimension, $FD_K$, the complexity is defined based on the deviation of the trajectory of the time series from the simplest path, i.e., the straight line between the first and last points (9).

$$FD_K = \frac{log(n)}{log(n) + log\left(\frac{d}{L}\right)} \quad (9)$$

where $d$ is the farthest distance among all those measured between the first point of the signal and all the others; $L$ is the total length of the waveform and is calculated as the sum of the distances between successive points of the signal; and $n = L/a$ is the number of units that compose the time series. In particular, to avoid different units leading to different fractal dimensions, a general unit, $a$, is defined as the average distance between subsequent points [34, 35].

Using Higuchi fractal dimension, $FD_H$, the complexity is assessed by analyzing how the pattern of the signal changes when sampled at different time intervals. F irst, starting from the time series $x = x(1), x(2), ..., x(N)$ of length $N$, $k$ new sequences are defined following (10), where $m = 1, 2, ..., k$ is the initial time value; $k \in \left[1, k_{max}\right]$ is the discrete time delay; and $\lfloor a \rfloor$ indicates the integer part of $a$. In this study, $k = 10$ [32, 35].

$$x_k^m = \left[x(m), x(m+k), x(m+2k), ..., x\left(m + \lfloor \tfrac{N-m}{k} \rfloor k\right)\right] \quad (10)$$

The length of each $x_k^m$ is evaluated, $L_m(k)$ (11), before averaging all those associated with the same delay, $k$ (12). Therefore, $FD_H$ is determined as the slope of the least squares linear best fit to the distribution of $L(k)$ versus $1/k$ on a double logarithmic scale.

$$L_m(k) = \tfrac{N-1}{k} \left[ \tfrac{1}{\lfloor \tfrac{N-m}{k} \rfloor} \sum_{i=1}^{\lfloor (N-m)/k \rfloor} |x(m+ik) - x(m+(i-1)k)| \right] \tfrac{1}{k} \quad (11)$$

$$L(k) = \sum_{m=1}^{k} L_m(k) \propto k^{-FD_H} \quad (12)$$

Similarly to $FD_K$, Petrosian fractal dimension, $FD_P$, quantifies the complexity of the time series in relation to how much it deviates from a straight line. In particular, given a signal of length $N$, Petrosian's definition (13) focuses on the number of sign-changes in its first derivative, thus emphasizing the rate of slope reversals [33, 34].

$$FD_P = \frac{log(N)}{log(N) + log\left(\frac{N}{N + 0.4N_\Delta}\right)} \quad (13)$$

## Frequency-domain features

In the list below we report mathematical details on how to compute each feature.

- **Spectral energy** represents the total energy of the time series and is defined by integrating the Power Spectral Density (PSD) of the signal.
  **Relative powers** of all the EEG frequency bands, i.e., **delta** ($\delta$, $0.5$–$4\ Hz$), **theta** ($\theta$, $4$–$8\ Hz$), **alpha** ($\alpha$, $8$–$12\ Hz$), **sigma** ($\sigma$, $12$–$16\ Hz$), **beta** ($\beta$, $16$–$30\ Hz$), and **gamma** ($\gamma$, $30$–$35\ Hz$) are also calculated. These estimate how the signal energy is distributed across its several frequency components.
  In addition, several **ratios between frequency bands** are included as features, i.e., $\delta/\theta$, $\delta/\sigma$, $\delta/\beta$, $\theta/\alpha$, $\delta/\alpha$, $\alpha/\beta$, $\delta/(\alpha + \beta)$, $\theta/(\alpha + \beta)$, $\delta/(\alpha + \beta + \theta)$.

- Unlike previous entropy metrics, **spectral entropy**, **SpecEn** (14), and **Renyi entropy**, **RenyiEn** (15) are measured in the frequency domain. In particular, they assess the complexity of the EEG signal by working on its normalized power spectrum [39].

$$SpecEn = -\sum_{f} p(f) \log[p(f)] \qquad (14)$$

$$RenyiEn = -\log\left[\sum_{f} p^{2}(f)\right] \qquad (15)$$

- **Spectral centroid** represents the center of mass of the spectrum of the time series and is calculated as the frequency-weighted mean of the PSD of the signal [40, 41].

- **Spectral flatness** is computed as the ratio of the geometric mean of the power spectrum to its arithmetic mean [40]. It evaluates how much noise-like the signal is [41].

- **Spectral spread** quantifies the dispersion of the power spectrum around its spectral centroid. Numerically, it is given by the weighted mean of the PSD of the signal and the weights are defined as the squared differences between the spectral centroid and all the examined frequencies [41].

- **Spectral crest factor** provides information about how extreme the peaks of the spectrum of the signal are. It is evaluated as the ratio of the maximum power spectrum to the mean PSD of the signal [42].

- **Spectral roll-off** is defined as the frequency beneath which a certain percentage of the overall energy lies. In this study, this feature is determined relative to 85% of the energy [43].

- The four **spectral central moments** in statistics, i.e., **mean**, **variance**, **skewness**, and **kurtosis**, provide insights into the shape and distribution of the PSD of the signal.

### Feature selection

The feature selection algorithm we choose relies on pairwise feature similarity, which is evaluated using the maximal information compression index (MICI). This metric has been shown to outperform two other commonly used feature similarity measures [44, 46].

Let Σ be the two-by-two covariance matrix of features x and y, $\lambda_2$ is defined as the smallest eigenvalue of Σ (16, 17).

$$\lambda_2(x, y) = \frac{1}{2}\left[var(x) + var(y) - \sqrt{A(x, y)}\right] \quad (16)$$

$$A(x, y) = [var(x) + var(y)]^2 - 4var(x)var(y)\left[1 - \rho(x, y)^2\right] \quad (17)$$

where $var$ stands for variance and ρ represents Pearson's correlation coefficient.

In particular, what is used here is a modified version of the MICI, as it gets normalized by the sum of the variances of the features to not have sensitivity to the features' scale (18) [46].

$$\lambda_{2, norm}(x, y) = \frac{\lambda_2(x,y)}{var(x) + var(y)} \quad (18)$$

The Feature Selection using Feature Similarity (FSFS) algorithm [44] is based on the k-nearest neighbors (kNN) principle, i.e., it divides the initial feature subset into homogeneous clusters, before selecting only the most representative feature from each such cluster. Iteratively, the algorithm selects only the feature showing the most compact subset, i.e., the lowest distance from its farthest neighbor, thus removing all the k-nearest ones. A constant error threshold, ε, is defined by the MICI value between the feature selected at the first iteration and its k-th neighbor. This is used to adapt the algorithm in such a way that at each iteration whenever the smallest $\lambda_2$ is lower than ε, the number of k-nearest neighbors (k) decreases [44].

The best initial value for k is set according to two metrics, i.e., the representation entropy ($H_R$) (19) and the redundancy rate ($RR$) (20). The former quantifies the information compression. Higher values are linked to a more balanced feature selection process, i.e., lower redundancy within the feature subset [44, 45].

$$H_R = -\sum_{i=1}^{N} \hat{\lambda}_i \log \hat{\lambda}_i \qquad (19)$$

where $\hat{\lambda}_i$ is the normalized eigenvalue of the covariance matrix of the feature subset of size N, having i=1, 2, ..., N and $\hat{\lambda}_i = \lambda_i / \sum_{i=1}^{N} \lambda_i$.

The redundancy rate evaluates the redundant information within the feature subset with larger values indicating a strong correlation among features [45, 47].

$$RR = \frac{1}{N(N-1)} \sum_{f_i, f_j \in F, i>j} \rho_{i,j} \qquad (20)$$

where $\rho$ is the Pearson's correlation coefficient measured for each pair of features ($f$) of the target subset ($F$) of size N. The best number of k-nearest neighbors is independently found for each sleep stage and for each pair of PSG and in-ear-EEG channels as the one that maximizes $H_R$. Hence, the chosen k-value is validated using the RR metric. First, we identify the k-value related to the maximum of the representation entropy, $H_{R,max}$ - once verified that the latter is greater than the reference value, i.e., the representation entropy evaluated on the initial feature subset, $H_{R,ref}$. Hence, we validate the chosen k-value, $k_0$, by verifying that the corresponding redundancy rate, $RR(k_0)$, is lower than the reference value, i.e., the redundancy rate evaluated on the initial feature subset, $RR_{ref}$. If either of these two conditions — $H_R(k_0) = H_{R,max} > H_{R,ref}$ and $RR(k_0) < RR_{ref}$ — is not met, the analysis is iterated considering the next highest $H_R$ value; and if both conditions are never simultaneously fulfilled, the initial number of k-nearest neighbors is set equal to zero.

# SUPPLEMENTARY TABLES

**Table S1.**

$Soft\text{-}Agreement$ values computed on each of the three scorers and for each subject on the PSG data source. For each subject, we report the most reliable scorer with the corresponding cell with edges highlighted in bold.

| | Soft-agreement for PSG scorers | | |
|---|---|---|---|
| | Scorer 1 | Scorer 2 | Scorer 3 |
| Subject 1 | 0.9366 | 0.9673 | **0.9836** |
| Subject 2 | 0.8840 | **0.9672** | 0.9540 |
| Subject 3 | 0.9443 | **0.9550** | 0.8737 |
| Subject 4 | 0.9588 | **0.9897** | 0.9670 |
| Subject 5 | 0.8261 | **0.9855** | 0.9524 |
| Subject 6 | 0.8950 | **0.9391** | 0.8466 |
| Subject 7 | 0.9582 | **0.9940** | 0.9681 |
| Subject 8 | 0.8929 | **0.9841** | 0.9167 |
| Subject 9 | **0.9791** | 0.9248 | 0.9687 |
| Subject 10 | 0.9561 | **0.9728** | 0.9477 |
| **Averaged** | 0.9231 ± 0.0447 | **0.9680 ± 0.0215** | 0.9379 ± 0.0427 |

**Table S2.**

*Soft-Agreement* values computed on each of three scorers and for each subject on the in-era-EEG data source. For each subject, we report the most reliable scorer with the corresponding cell with edges highlighted in bold.

| | Soft-agreement for in-ear-EEG scorers | | |
|---|---|---|---|
| | Scorer 1 | Scorer 2 | Scorer 3 |
| Subject 1 | 0.8773 | 0.9121 | **0.9550** |
| Subject 2 | 0.8709 | 0.8840 | **0.9540** |
| Subject 3 | 0.8522 | 0.8244 | **0.9465** |
| Subject 4 | 0.8784 | 0.8247 | **0.9443** |
| Subject 5 | 0.8468 | 0.9006 | **0.9379** |
| Subject 6 | 0.7479 | 0.8382 | **0.9454** |
| Subject 7 | **0.9821** | 0.8825 | 0.9203 |
| Subject 8 | **0.8651** | 0.7897 | 0.8452 |
| Subject 9 | 0.8225 | **0.9353** | 0.9165 |
| Subject 10 | 0.7762 | **0.9519** | 0.8891 |
| **Averaged** | 0.8519 $\pm$ 0.0602 | 0.8743 $\pm$ 0.0504 | **0.9254 $\pm$ 0.033** |

**Table S3.**

List of all the extracted time-domain features. Abbreviations: Detrended Fluctuation Analysis (DFA); Singular Value Decomposition (SVD). **\*** features depending on amplitude, thus computed on normalized signals.

| Time-domain features | | |
|---|---|---|
| Standard deviation* | DFA exponent | Hjorth activity* |
| Skewness | Approximate entropy | Hjorth mobility |
| Kurtosis | Sample entropy | Hjorth complexity |
| Maximum first derivative* | SVD entropy | Katz fractal dimension |
| Interquartile range* | Permutation entropy | Higuchi fractal dimension |
| Number of zero-crossings | Lempel-Ziv complexity | Petrosian fractal dimension |

**Table S4.**

List of all the extracted frequency-domain features. **\*** features depending on amplitude, thus computed on normalized signals.

| Frequency-domain features | | |
|---|---|---|
| Spectral energy* | $\delta/\theta$ power ratio | Spectral centroid |
| Relative $\delta$ power band | $\delta/\sigma$ power ratio | Spectral crest factor |
| Relative $\theta$ power band | $\delta/\beta$ power ratio | Spectral flatness |
| Relative $\alpha$ power band | $\delta/\alpha$ power ratio | Spectral roll-off |
| Relative $\sigma$ power band | $\theta/\alpha$ power ratio | Spectral spread |
| Relative $\beta$ power band | $\alpha/\beta$ power ratio | Spectral mean* |
| Relative $\gamma$ power band | $\delta/(\alpha + \beta)$ power ratio | Spectral variance* |
| Spectral entropy | $\theta/(\alpha + \beta)$ power ratio | Spectral skewness |
| Renyi entropy | $\delta/(\alpha + \beta + \theta)$ power ratio | Spectral kurtosis |

# FIGURE CAPTIONS

**Figure S1.** Raw in-ear-EEG 30-second data sample for each subject.

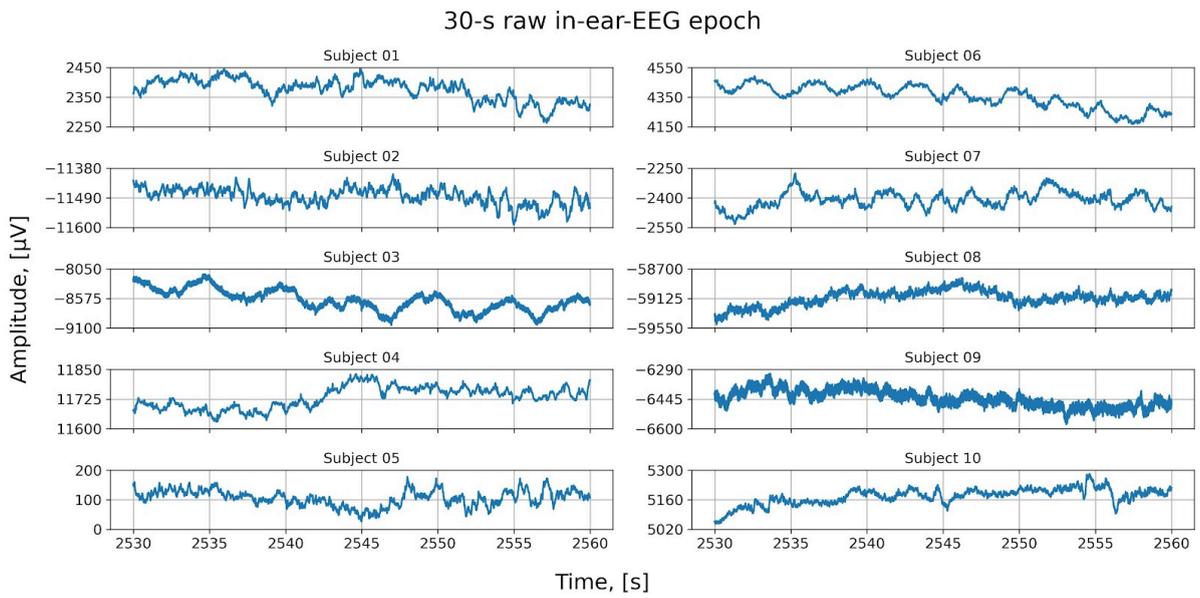

**Figure S2.** Pre-processed in-ear-EEG 30-second data sample for each subject.

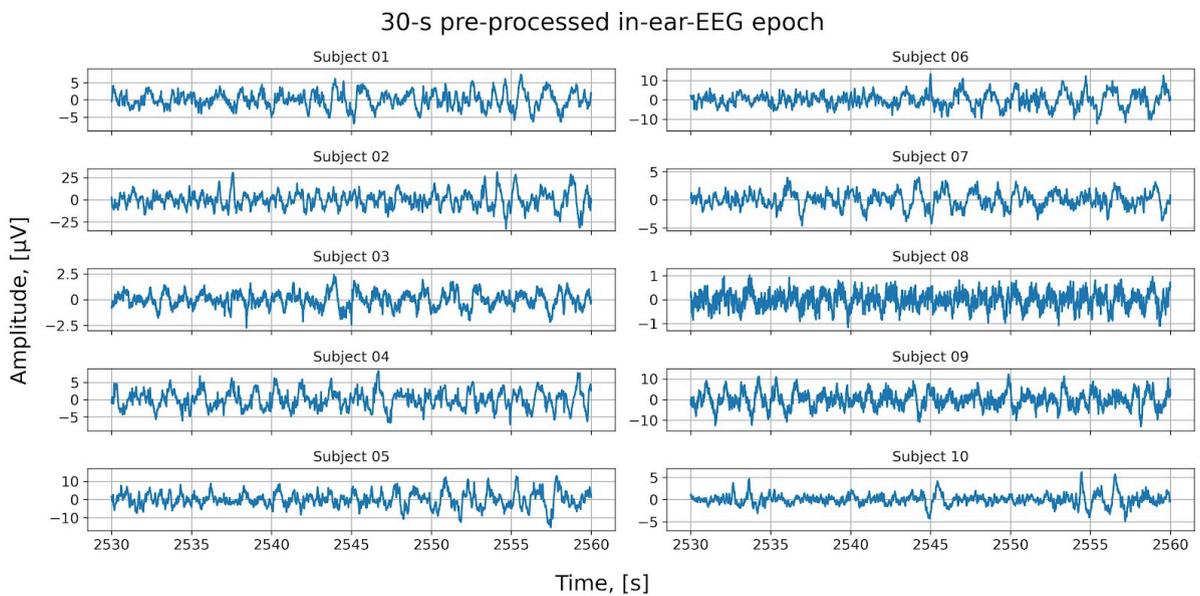

**Figure S3.** JSD-FSI similarity-scores distributions, i.e., distributions derived from Scalp-EEG-to-Scalp-EEG (blue line), and Scalp-EEG-to-In-ear-EEG (orange line) comparisons - for each subject in the awake stage. The area under the Scalp-EEG-to-In-ear-EEG distribution - related to JSD-FSI values greater than the minimum Scalp-EEG-to-Scalp-EEG score - is highlighted in purple.

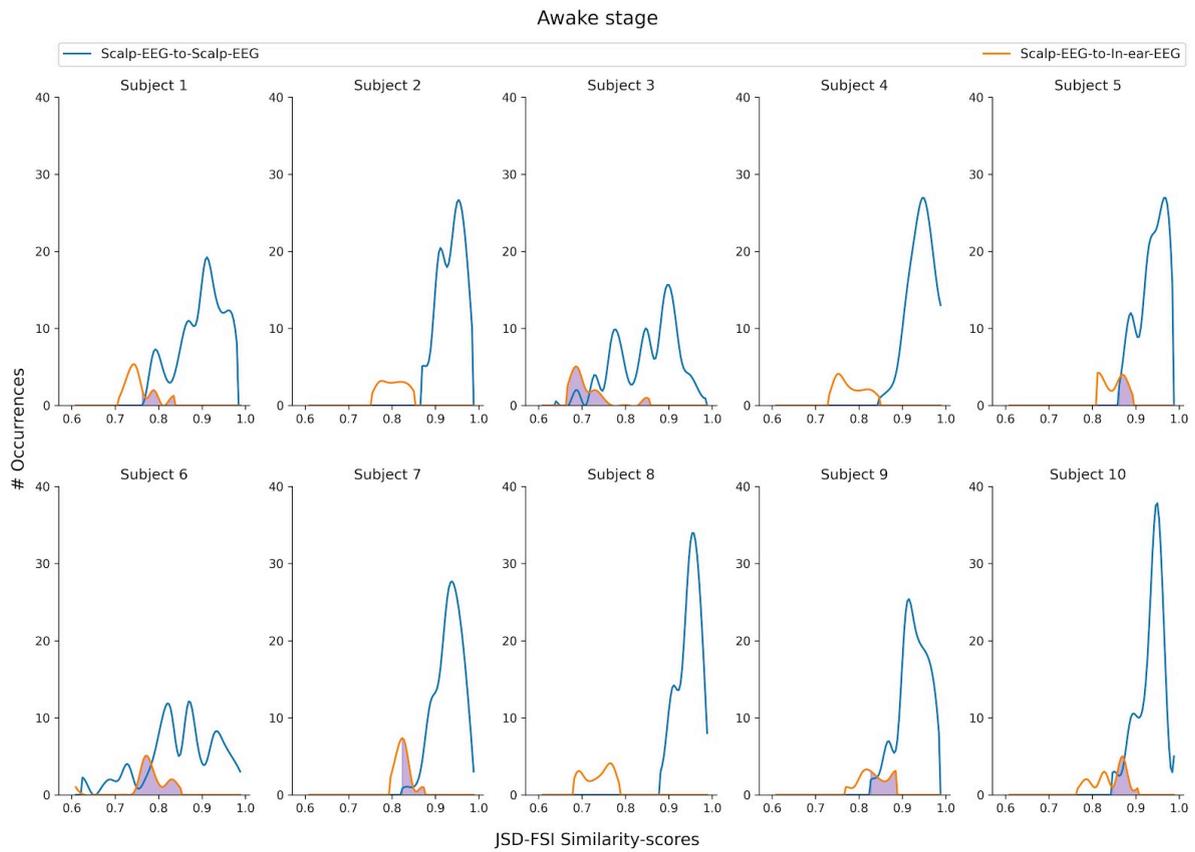

**Figure S4.** JSD-FSI similarity-scores distributions, i.e., distributions derived from Scalp-EEG-to-Scalp-EEG (blue line), and Scalp-EEG-to-In-ear-EEG (orange line) comparisons - for each subject in the NREM stage. The area under the Scalp-EEG-to-In-ear-EEG distribution - related to JSD-FSI values greater than the minimum Scalp-EEG-to-Scalp-EEG score - is highlighted in purple.

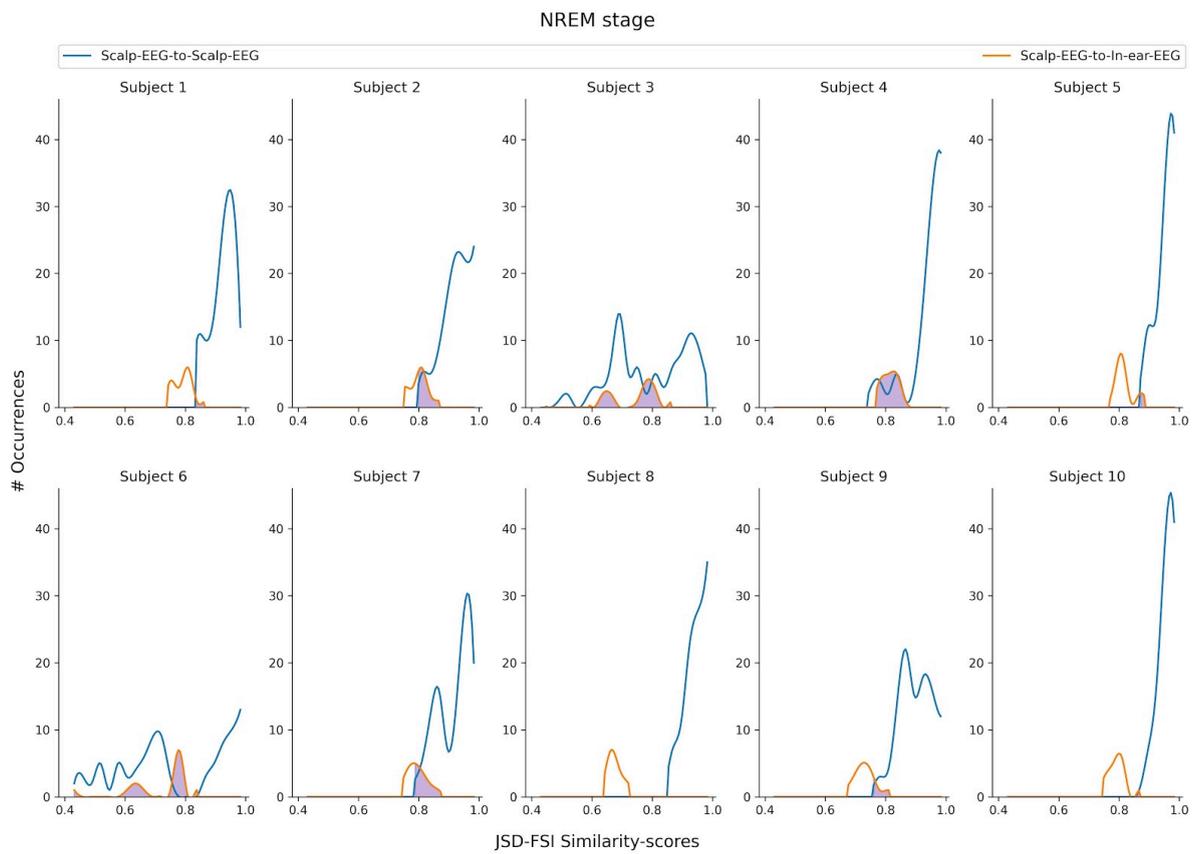

**Figure S5.** JSD-FSI similarity-scores distributions, i.e., distributions derived from Scalp-EEG-to-Scalp-EEG (blue line), and Scalp-EEG-to-In-ear-EEG (orange line) comparisons - for each subject in the REM stage. The area under the Scalp-EEG-to-In-ear-EEG distribution - related to JSD-FSI values greater than the minimum Scalp-EEG-to-Scalp-EEG score - is highlighted in purple. No JSD-FSI similarity-scores are reported for subjects 3 and 8.

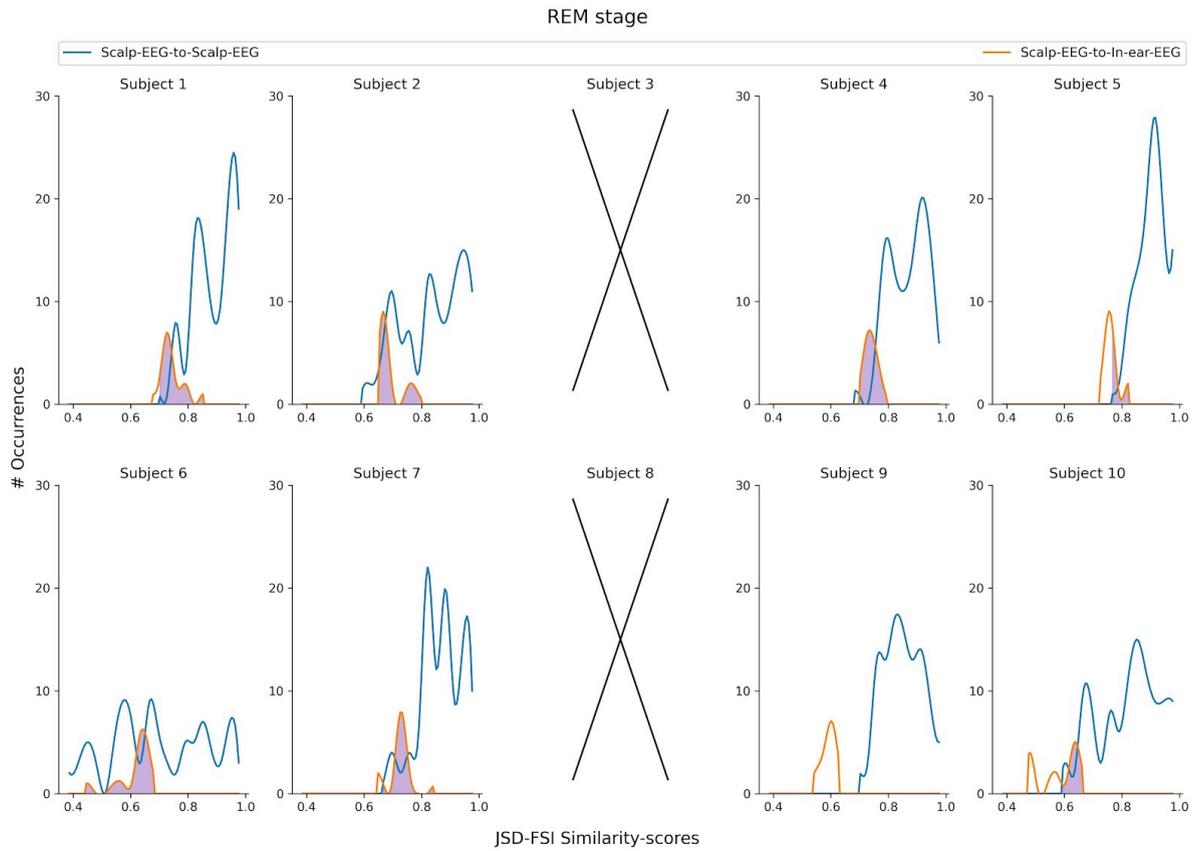

**Figure S6.** JSD-FSI similarity-scores distributions, i.e., distributions derived from EOG-to-EOG (green line), and EOG-to-In-ear-EEG (red line) comparisons - for each subject in the awake stage. The area under the EOG-to-In-ear-EEG distribution - related to JSD-FSI values greater than the minimum EOG-to-EOG score - is highlighted in purple.

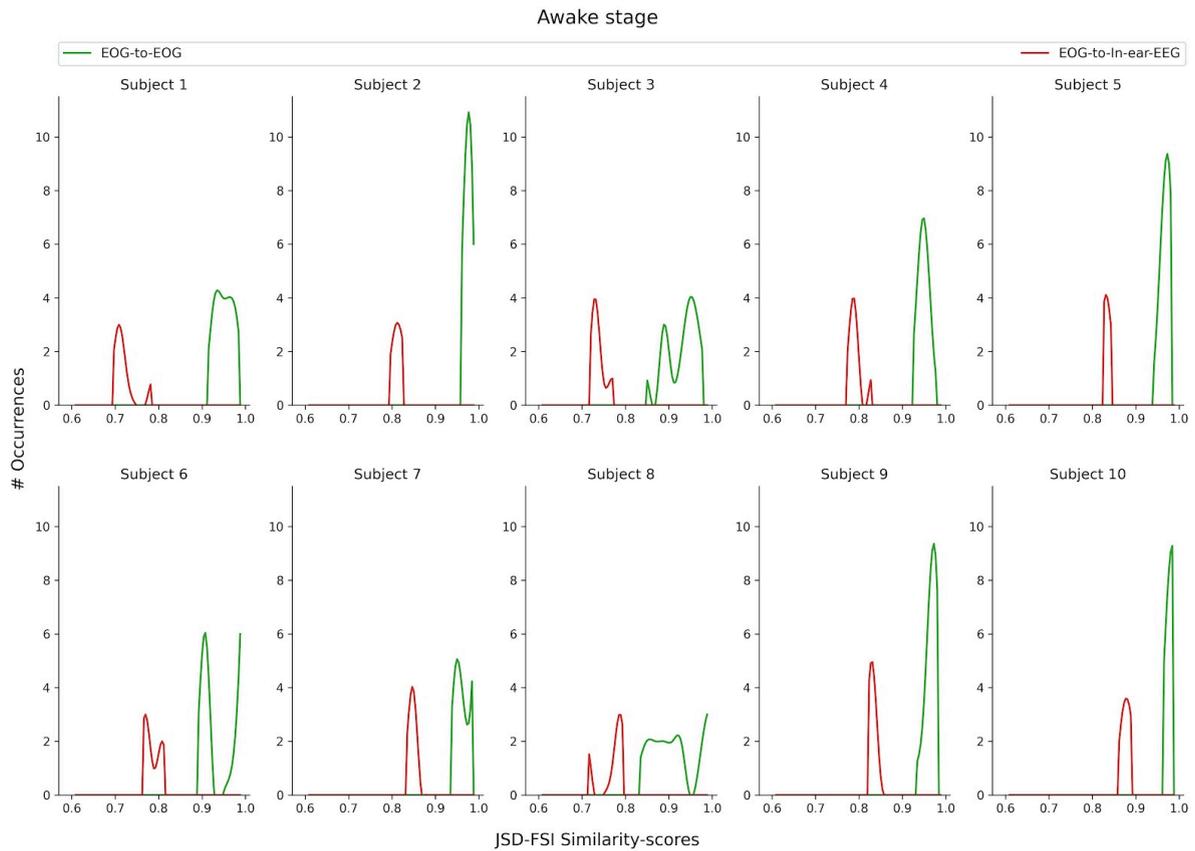

**Figure S7.** JSD-FSI similarity-scores distributions, i.e., distributions derived from EOG-to-EOG (green line), and EOG-to-In-ear-EEG (red line) comparisons - for each subject in the NREM stage. The area under the EOG-to-In-ear-EEG distribution - related to JSD-FSI values greater than the minimum EOG-to-EOG score - is highlighted in purple.

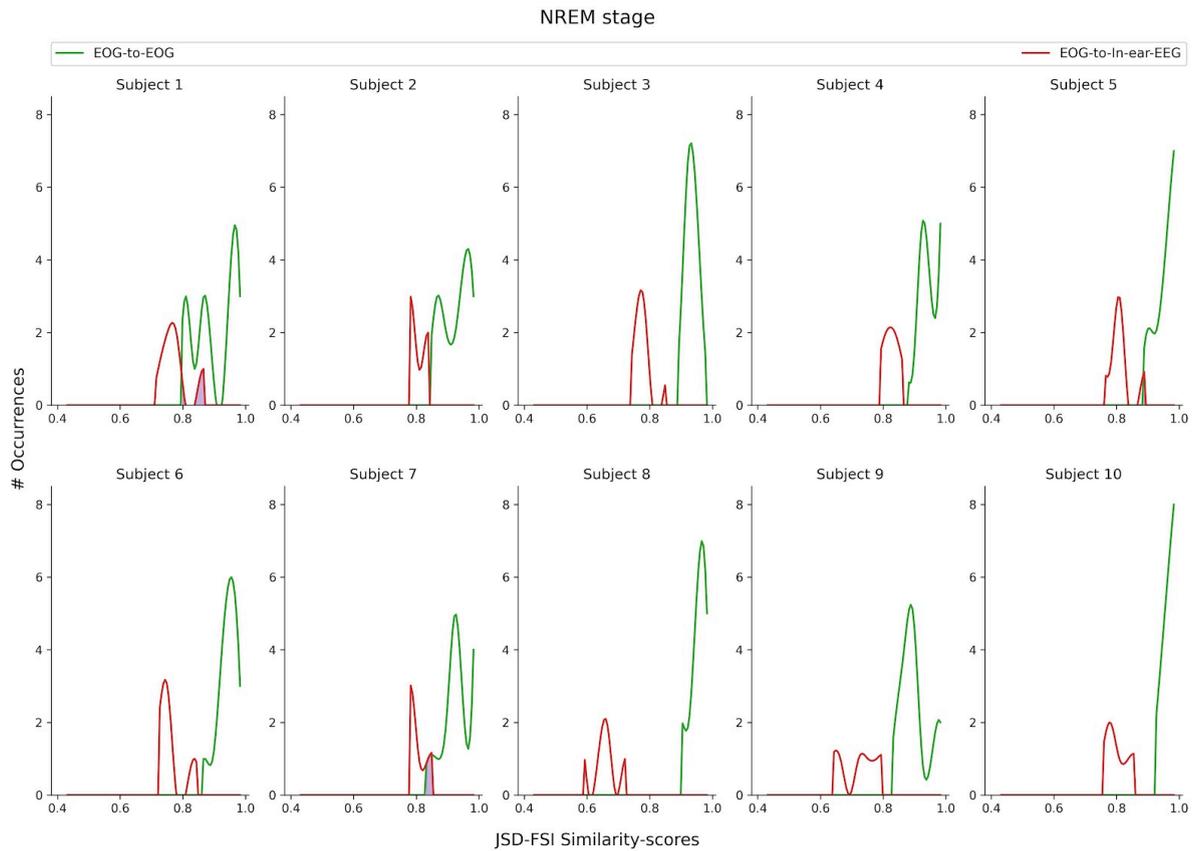

**Figure S8.** JSD-FSI similarity-scores distributions, i.e., distributions derived from EOG-to-EOG (green line), and EOG-to-In-ear-EEG (red line) comparisons - for each subject in the REM stage. The area under the EOG-to-In-ear-EEG distribution - related to JSD-FSI values greater than the minimum EOG-to-EOG score - is highlighted in purple. No JSD-FSI similarity-scores are reported for subjects 3 and 8.

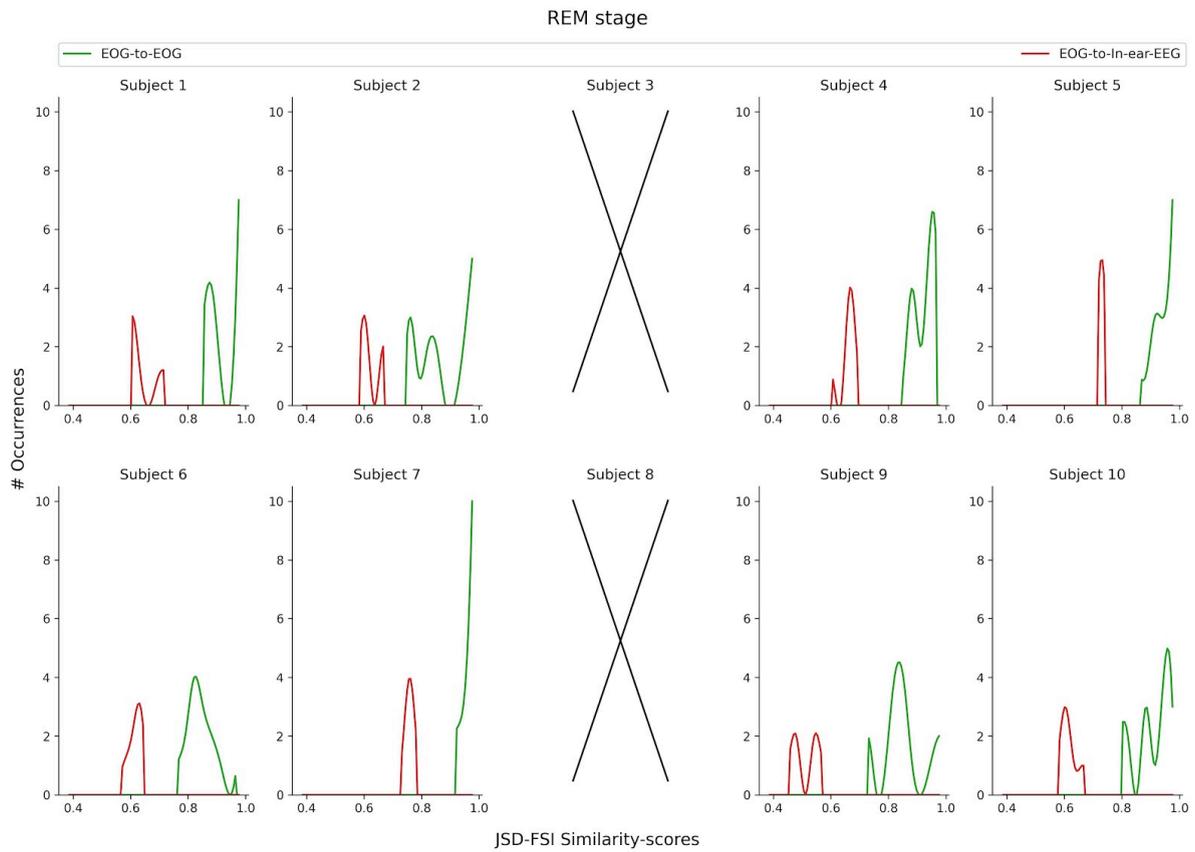